\documentclass[11pt,authoryear]{elsarticle}
\usepackage{amsmath}
\usepackage{amssymb}
\usepackage{graphicx,float}
\usepackage{natbib}
\usepackage[utf8]{inputenc}
\usepackage{booktabs}
\usepackage{array,multirow}
\usepackage{gensymb}
\usepackage{tabularray}
\usepackage{float}
\usepackage{graphicx}
\usepackage{subcaption}

\begin{document}

\begin{frontmatter}

\title{Threshold Resource Redistribution in Spatially-Structured Kinship Networks }
\author{Alina Kochocki }
\affiliation{organization={Department of Physics and Astronomy, Michigan State University}, 
city={East Lansing, MI}, 
postcode={48823}, 
country={USA}}
\ead{kochocki@msu.edu}

\begin{abstract}
We present a model for a threshold-based resource redistribution process in a spatially-explicit population, characterizing the relation between kinship network structure, local interactions and persistence. We find that population survival becomes possible for lower resource densities, but leads to increased network heterogeneity and locally centralized clusters. We interpret this in relation to a feedback between the kinship network structure and reproduction ability. Agents receive stochastic resources and solicit additional resources from connected individuals when below a minimum, with each agent contributing a fraction of their excess based on relatedness. We first analyze a fully-connected population with uniform redistribution fraction and discuss mean field expectations as well as finite size corrections. We extend this model to a hub-and-spoke network, exploring the impact of network asymmetry or centrality on resource distribution. We then develop a spatially-limited population model with diffusion, local pairing, reproduction and mortality. Redistribution is introduced as a function of relatedness (generational distance through most-recent common ancestor) and distance. Redistribution-dependent populations exhibit a higher level of relational closeness with increased clustering for agents of highest node strength. These results highlight the interaction of resource density, cooperation and kinship in a spatially-limited regime.
\end{abstract}


\end{frontmatter}

\section{Introduction}

\subsection{Resource Redistribution within Kinship Networks}

Resource redistribution occurs commonly within ecology, anthropology, sociology and economics \citep{Mauss1925Gift, Dragulescu_2000, repec:hdr:hdocpa:hdocpa-2007-25, Dolfing19, Beltran2023ReciprocityReview}. Wealth or energy transfer between agents follows some sharing rules or constraints. The result may benefit individual agents or the system as a whole \citep{Chappin2010, Carvalho2012FairSharing, Voorn2020, NDavis2022}. In this work, we focus on how resource redistribution acts to reduce variance between members of a population, increasing agent survival. While this could be described as a form of cooperation, agents are non-autonomous, and follow global redistribution rules. This system has general biological or ecological relevance, but also resembles wealth and resource transfer within kinship structures. 

We treat redistribution with respect to a value threshold. Individual agents are each initially provided with some sampled resource amount following a Poisson distribution of mean, $\mu$. Agents desire to acquire a resource minimum or threshold through cooperative requests. Each agent-pair can transfer a fraction, $\rho$, of excess resources –– only agents with excess resources will redistribute. These transfers generally lift the number of surviving agents, with efficacy dependent both on the transfer fraction and on $\mu$. 

This model of resource redistribution is relevant in systems where immediate, individual need is perceived, but global resource mean and variance is poorly known. General support systems may be ordered and heterogeneous, with biases in favor of certain population members. Within biological populations, agents are limited both spatially and temporally, and localization plays a substantial role in transport. Scaling the overall resource transfer amount acts to change the scale of individual agent dependence on the complimentary population – a few large, localized transfers as opposed to many small-value, diffuse transfers. 

We consider this process first for a fully connected population, providing mean field expectations and the impact of finite population size. We then consider a structured hub-and-spoke network, demonstrating the role of varied agent node strength. Finally, we model a synthetic spatially-limited kinship network, and explore population survival, extinction and structure as a function of cooperative redistribution and resource density. An agent's redistribution supply, survival and reproduction is made dependent on the size and proximity of its familial network, ultimately changing the expected network topology and diversifying the adaptive, dynamic behavior of the structure. We characterize the impact of redistribution on this system qualitatively and through numerical simulation. We conclude with a brief discussion of future paths for analytical descriptions. 

\section{Model and Methods}

\subsection{Cooperation Within a Fully Connected Population}

\subsubsection{Threshold Resource Redistribution }

We first describe a resource transfer between members of a population with a standardized rule for sharing. Within this toy model, each $i$th agent begins with an initial resource amount, $r_{i}$, sampled from, 
\begin{equation}
    P(X = r_{i}) = \dfrac{e^{-\mu} \mu^{r_{i}} }{ r_{i} !}.
\end{equation}
If each agent requires at least $r_{i} \geq \phi$, an agent's demand is $d_{i} = (\phi - r_{i})_{+}$. A complimentary $j$th agent has an excess $s_{j} = (r_{j} - \phi)_{+}$. A fraction, $\rho$, of excess resources may be transferred. Assuming agents do not request more than what is required, the transferred amount in a given interaction is, 
\begin{equation}
    \Delta = \textrm{ min} (\rho s_{j} , d_{i}).
\end{equation}
This agent-to-agent truncation impacts the resulting resource flow and must be modeled explicitly for the case of finite N. 

\subsubsection{Mean Field Treatment}

In the limit, $N \rightarrow \infty$, our expected total resource demand for $N$ (large) agents is, 
\begin{align}
D(\mu)
&= N \int_{0}^{\phi} (\phi - r)\, P(r)\, dr, \\
&= N \sum_{k=0}^{\lfloor \phi \rfloor}
    (\phi - k)\, \frac{e^{-\mu} \mu^k }{k!}, \\
&= N e^{-\mu},
\qquad \phi = 1 .
\end{align}
Here, we have also evaluated the result for the specific value of $\phi = 1$, which is used for demonstration later in the text. The total supply (shareable excess) within the population is, 
\begin{align}
S(\mu,\rho)
&= N\rho \int_{\phi}^{\infty} (r - \phi)\, P(r)\, dr, \\
&= N\rho \sum_{k=\lfloor \phi \rfloor + 1}^{\infty}
    (k - \phi)\, \frac{e^{-\mu} \mu^k }{k!}, \\
&= N\rho \left(\mu - 1 + e^{-\mu}\right),
\qquad \phi = 1 .
\end{align}
For survival of the deficit agent population, we must have at least $S(\mu, \rho) \geq D(\mu)$.

We can also define the expected, average demand with respect to the population, 
\begin{equation}
    \mathbb{E}[ (\phi - r )_{+} ] = \int_{0}^{\phi} (\phi - r) P(r) dr  = \dfrac{D(\mu)}{N}.
\end{equation}
Similarly, 
\begin{equation}
    \mathbb{E}[ (r - \phi)_{+} ] = \dfrac{S(\mu, \rho)}{N \rho}.
\end{equation}
As the Poisson mean expectation is $\mu$, we also have, 
\begin{equation}
    \mathbb{E}[ (r - \phi)_{+} ] = \mu - \phi + \mathbb{E}[ (\phi - r )_{+} ].
\end{equation}
The expected resource excess is determined by the mean resource density relative to the threshold and expected demand. 

We can derive a condition for the transfer fraction, $\rho$, such that all agents survive. Expected supply exceeds expected demand, 
\begin{align}
    \rho \mathbb{E}[ (r - \phi)_{+} ] &\geq \mathbb{E}[ (\phi - r )_{+} ], \\
    \rho (  \mu - \phi + \mathbb{E}[ (\phi - r )_{+} ] ) &\geq \mathbb{E}[ (\phi - r )_{+} ], \\
    \rho (  \mu - \phi + D/N ) &\geq D/N.
\end{align}
We then have the general condition, 
\begin{equation}
    \rho( \mu - \phi) \geq (1 - \rho) \mathbb{E}[ (\phi - r )_{+} ].
\end{equation}
A phase transition appears where all members of the population may survive, dependent on $\rho$, $\mu$ and $\phi$. Full survival occurs if and only if $\mu \geq \mu_{c}(\phi, \rho)$. The location of the transition, $\mu_{c}$, is such that, 
\begin{equation}
\mu_c(\phi,\rho) =
\begin{cases}
\phi, & \rho = 1, \\[4pt]
>\phi, & 0 < \rho < 1.
\end{cases}
\end{equation}

Finally, the surviving deficit population fraction in the mean field limit is given by, 
\begin{equation}
    f = \textrm{min}\Bigg(1,  \dfrac{S}{D} \Bigg) = \textrm{min}\Bigg(1, \dfrac{\rho \mathbb{E}[ (r - \phi)_{+} ] }{ \mathbb{E}[ (\phi - r )_{+} ] } \Bigg) .
\end{equation}
This result is relevant only for the case of large $N$, reflecting mean supply and demand. 

\subsubsection{Survival Fraction as a Function of Transfer Fraction}

\begin{figure}
    \centering
    \includegraphics[width=0.6\linewidth]{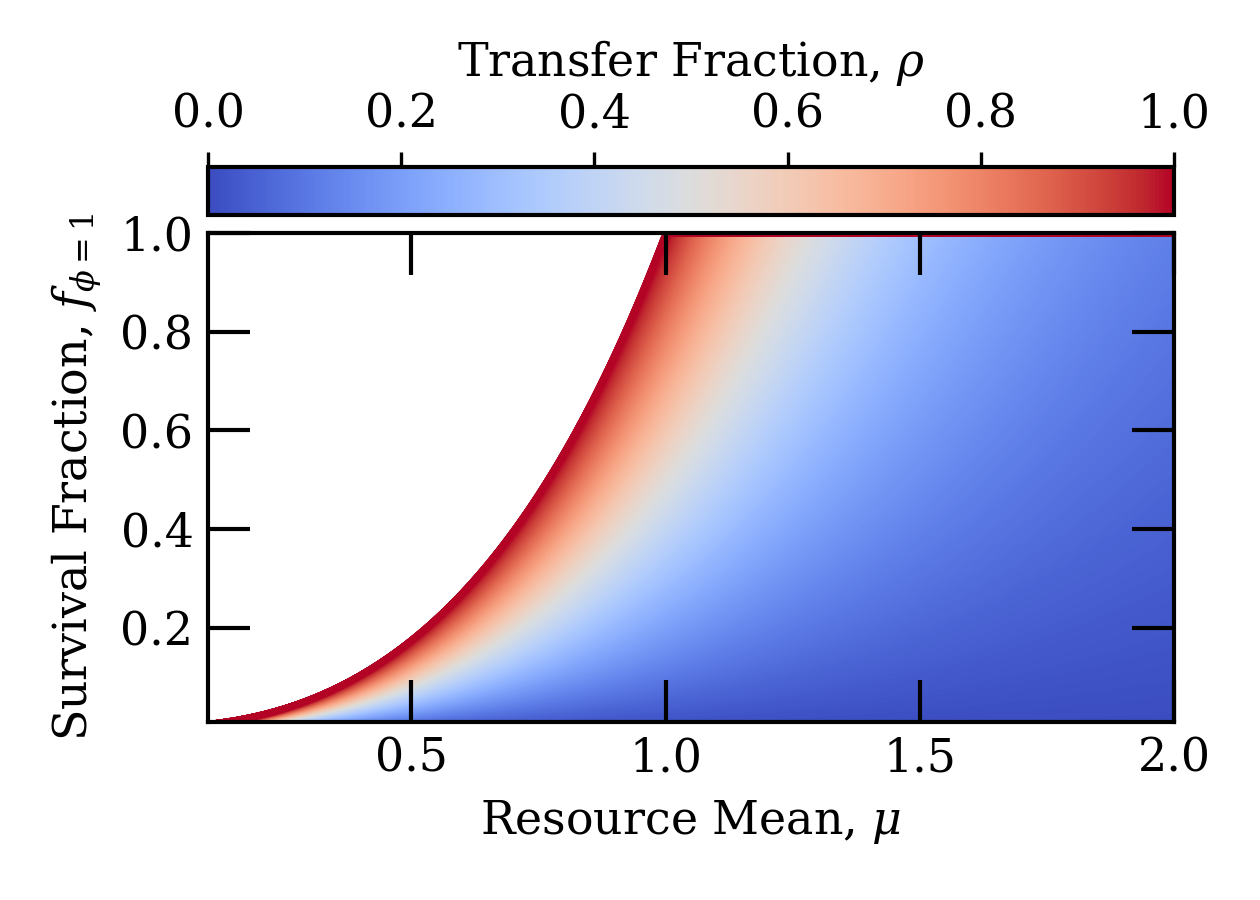}
    \caption{Mean field survival fraction as a function of initial resource distribution mean and transfer fraction. We provide a simple demonstration of the expected mean field behavior for this threshold redistribution process. The surviving fraction is relative to the initial number of deficit agents. A threshold of one is assumed. If the resource mean is less than one, increased sharing is insufficient for full population survival. A larger resource mean requires a lower transfer fraction to effectively eliminate remaining downwards Poisson fluctuations. }
    \label{fig:mf_fully_con}
\end{figure}

In the case that $\phi = 1$, we have the mean field expectation, 
\begin{equation}
    f|_{\phi = 1} = \textrm{min}\Bigg(1, \dfrac{\rho ( \mu - 1 + e^{-\mu}) }{ e^{- \mu}} \Bigg).
\end{equation}
Here, the surviving fraction is a function of the mean resource density, $\mu$, and transfer fraction, $\rho$. This represents survival of the initial deficit population with insufficient resources. We provide a demonstration of $f|_{\phi = 1}$ as a function of $\rho$ and $\mu$ in Figure \ref{fig:mf_fully_con}.

\subsubsection{Finite Size Corrections }

In the case of a finite population, our survival condition is smoothed due to stochastic fluctuations. Given the total supply and demand observed from $N$ agents, $D_{N}$ and $S_{N}$, we define the variances, 
\begin{align}
    \textrm{Var}(D_{N}) &= N \sigma_{D}^{2} = N \textrm{ Var}( (\phi - r)_{+} ), \\
    \textrm{Var}(S_{N}) &= N \sigma_{S}^{2} = N \rho^{2} \textrm{ Var}( (r - \phi)_{+} ).
\end{align}

By the central limit theorem \citep{kallenberg2002foundations}, the observed supply and demand are approximately Gaussian for large $N$. The probability of full survival is,
\begin{equation}
\mathcal{P}(N) = \Phi \left( \frac{\sqrt{N}(S-D)}{\sqrt{\sigma_S^2 + \sigma_D^2}} \right).
\end{equation}
The corresponding expected survival fraction is, $\mathbb{E}[f_N] = S/D + O(1/N)$. Further discussion and derivation is provided in Appendix~A. 

\subsubsection{Finite Size and Truncation Demonstration}
\begin{figure}[t]
    \centering
    \includegraphics[width=\linewidth]{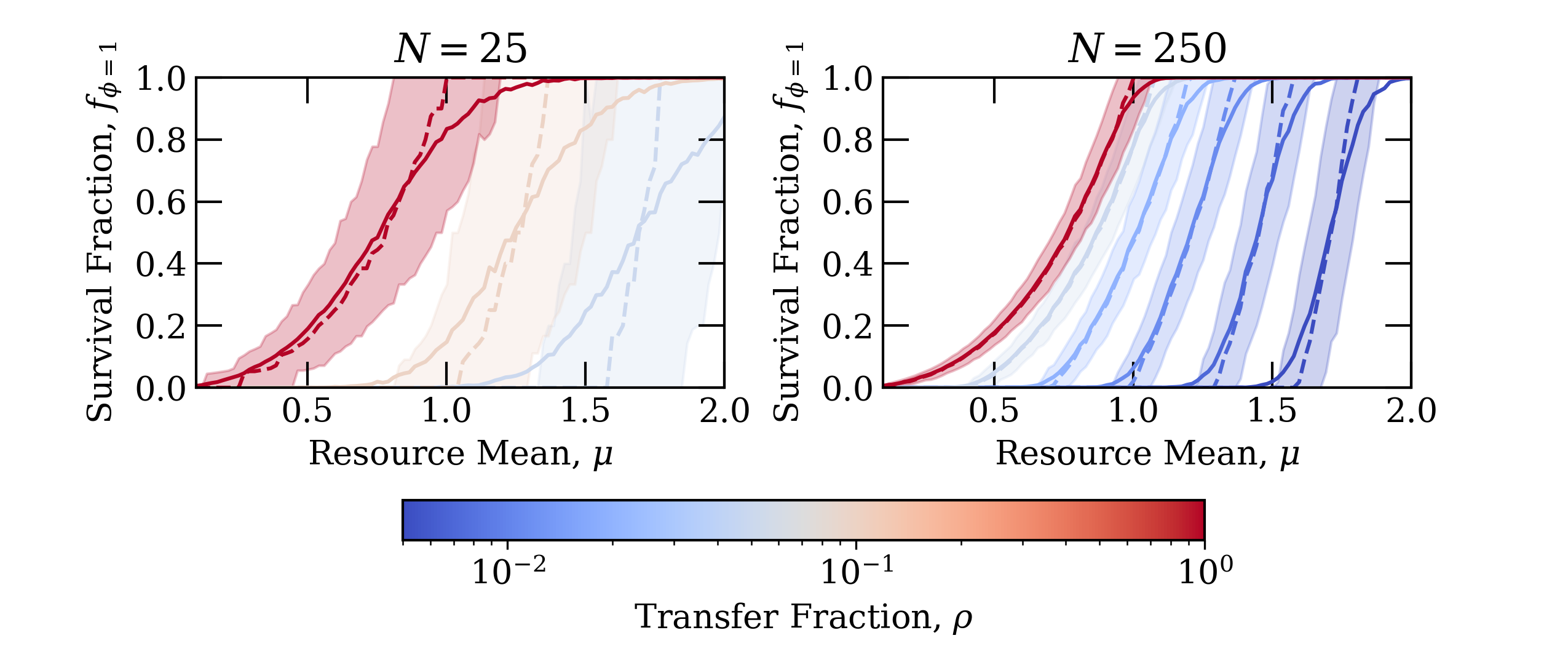}
    \caption{Survival fraction as a function of initial resource distribution mean and transfer fraction. We demonstrate the behavior of our threshold redistribution process for networks of 25 and 250 agents. A threshold of $\phi = 1$ is assumed. Results of these numerical simulations are plotted as a function of resource density for $\rho = 0.05, 0.1$ and 1.0 on the left. Additional, selected values are plotted on the right. With larger $N$, there are additional opportunities for resource transfer, leading to a higher survival fraction for the same $\rho$ relative to the low-$N$ case. The shaded regions represent the 68$\%$ central containment percentiles of the resulting survival fractions of $10^{4}$ simulations. The mean (true to the colorbar) is plotted with a solid line, while the median is also shown with a dashed line to indicate skewness. }
    \label{fig:mf_fully_con_dem}
\end{figure}

We can demonstrate the expected variance for a threshold of $\phi = 1$ with finite $N$. We consider networks of $N = 25$ and 250 nodes. Deficit agent survival fraction is determined as a function of $\rho$ and $\mu$. With each sampled value of $\mu$, the 68$\%$ central containment region of the result of $10^{3}$ numerical simulations is presented. These results are shown in Figure \ref{fig:mf_fully_con_dem}. 

With a finite population size, variance in the total supply and demand become relevant. The initial sampled resource amount may be insufficient to satisfy all agents, even if the expected resource supply is in excess of the expected resource demand. These fluctuations contribute to variance in the survival fraction. 

As $N$ increases, total supply and demand become increasingly self-averaging for a given $\mu$ and $\rho$. Survival is increased by the larger number of independent agents, more effectively diminishing extreme fluctuations and reducing variance in the remaining agent demand. A maximal survival fraction for the same $\mu$ as a function of $\rho$ is achieved more quickly relative to the low-$N$ case. 

Similar to the probability of full survival, the variance of the survival fraction suggested by the 68$\%$ central containment region decreases with $N$. As the survival fraction is maximally one, skewness in these survival fraction distributions becomes notable. 

\subsection{Cooperation Within a Centralized Network}

We also consider cooperation or resource distribution over a structured network. As a toy example, we treat a hub-and-spoke network, in which a central hub is connected to $N - 1$ spokes. Each spoke is connected only to the central hub. The structure of the network leads to the preferential survival of the hub, which is also true for the nodes of general network structures with higher relative node strength. We again assume a total population of size, $N$. The same resource transfer fraction, $\rho$, and resource distribution, $P(X)$, is assumed, along with the desired resource threshold, $\phi$. 

\subsubsection{Heterogeneous Mean Field Case}

We first consider large $N$, treating the expected behavior for the spoke and hubs separately. Of the population, a fraction $1 - 1/N$ of agents correspond to spokes, each with a degree of one. The mean initial demand and supply expected of each spoke is, 
\begin{align}
    D_{*} &= \mathbb{E}[(\phi - r)_{+}] = \sum_{k = 0}^{\phi - 1} (\phi - k) P(X = k), \\
    S_{*} &= \rho \mathbb{E}[(r - \phi )_{+}] = \rho \sum^{\infty}_{k = \phi + 1} (k - \phi) P(X = k).
\end{align}
A fraction $1/N$ of the population represents the hub. This agent has degree $N - 1$ and the same initial mean supply and demand as an individual spoke. We define the initial, observed hub supply and demand as $D_{h}$ and $S_{h}$.

Owing to the differences in node degree, there is an asymmetry in resource transfer across the network. The hub can request from any spoke, while all spokes rely on the initial resource supply of the hub. As the hub will only accept resources until its own threshold is met, redistribution is bottlenecked through this single node. With large $N$, the hub is generally satisfied while resource variance remains within the spokes. We can also define the total spoke demand and supply, $D_{s}^{\textrm{tot}}$ and $S_{s}^{\textrm{tot}}$, 
\begin{align}
    D_{s}^{\textrm{tot}} &= (N - 1) D_{*}, \\
    S_{s}^{\textrm{tot}} &= (N - 1) S_{*}. 
\end{align}
As hub survival is contingent on spoke supply, the condition for hub survival is, 
\begin{equation}
    S_{s}^{\textrm{tot}} \geq D_{h}.
\end{equation}
Such a scenario becomes feasible only when resources of the hub are allowed to vary with respect to the large-$N$ spoke field expectation. Similarly, the condition for total spoke survival is the existence of a sufficient hub excess, 
\begin{equation}
    S_{h} \geq D_{s}^{\textrm{tot}}. 
\end{equation}
Otherwise, only a fraction of deficit nodes, $f_{s}$, can be satisfied,
\begin{equation}
    f_{s} = \textrm{min}\bigg( 1, \dfrac{ S_{h}}{ D_{s}^{\textrm{tot}}} \bigg).
\end{equation}

Survival of all spokes and the hub requires minimal demand from the spokes, $\mu \gg \phi$. Survival of only the hub and potentially a slightly lifted fraction of spokes is more likely, or generally expected dependent on $\mu$ and $\rho$. In the case of a Poisson distribution with $\phi = 1$, the mean-field spoke survival fraction in the partial survival regime is, 
\begin{equation}
    f_{s}|_{\phi = 1} = \dfrac{ \rho (\mu - 1 + e^{-\mu})}{(N - 1)e^{-\mu}}.
\end{equation}
This is scaled down by a factor of $\sim$$N$ relative to the fully connected case, owing to the network asymmetry. As $N \rightarrow \infty$, spoke survival is negligible owing to the finite supply. The corresponding mean-field hub survival condition is, 
\begin{equation}
    \dfrac{\rho (N - 1) (\mu - 1 + e^{-\mu}) }{ e^{-\mu}} \geq 1.
\end{equation}
This is scaled up by a factor of $\sim$$N$ relative to the fully connected case, again owing to the network structure. In the case, $N \rightarrow \infty$, the hub is essentially guaranteed to survive given any $\rho > 0$.

\subsubsection{Finite Size Corrections }

Given a finite population of size $N$, variance in the total expected spoke supply and demand becomes important within our heterogeneous network. If a single spoke or hub has demand $d_{i} = (\phi - r_{i})_{+}$, we define $\mathbb{E}[d] = D_{*}$ and $\textrm{Var(d)} = \sigma^{2}_{D}$. Similarly, a node supply of $s_{i} = (r_{i} - \phi )_{+}$ has $\mathbb{E}[s] = S_{*}$ and $\textrm{Var(s)} = \sigma^{2}_{S}$. Following the central limit theorem, the probability of hub survival is then, 
\begin{equation}
    \mathcal{P }_{h}(N)  = \Phi \left( \dfrac{ (N - 1)S_{*}  - D_{*}}{ \sqrt{ (N - 1)\sigma_{S}^{2} + \sigma_{D}^{2}} } \right).
\end{equation}
The mean condition reduces to the mean field expectation. The width of this survival transition is $\sim$$1/\sqrt{N}$. The variance in hub survival for a given resource expectation, $\mu$, reduces with the number of spokes. As $N$ becomes large, the survival probability tends towards one. A longer discussion is provided within Appendix~B.

The probability of individual spoke survival is such that, 
\begin{equation}
    \mathcal{P}_{s}(N) = \Phi \left( \dfrac{S_{*}/(N - 1) - D_{*}}{ \sqrt{ \sigma_{S}^{2}/(N - 1) + \sigma_{D}^{2}}} \right).
\end{equation}
As the available supply of the hub is potentially shared across $N - 1$ spokes, the expected supply mean and variance in a single exchange is equivalent to the initial supply and variance divided across this population size. This is conditional on hub survival. As $N$ becomes large, the survival probability of deficit spoke agents tends towards zero. 

\subsection{Cooperation over a Spatially-Explicit Kinship Network }

Cooperation within kinship or familial networks has been widely documented and discussed \citep{Hamilton1964, Hamilton1996, Dawkins1976, Frank1998}. This may occur as explicit resource or wealth redistribution, but may also take the form of service acts, information transfer or other commonly overlooked labor. There are likely both biological influences for this preferential treatment (e.g. instinctual parental-child bonds) as well as cultural or apparently learned behavior \citep{faria2020does, LehmannRousset2010, richerson2016cultural, mitchell2021ontogeny}. While cooperation with more distant relations is common, in this work we explore a scenario in which kinship networks are most relevant, representing the leading order for resource transfer. 

Previous work has characterized cooperative social structures in early human societies, finding that group size was closely linked to resource density and volatility \citep{kaplan2009evolutionary, ayala2018hunter, dyble2016networks}. Larger cooperative groups were feasible and preferable when the resource density mean was sufficient for survival but the distribution was highly variable. Simple redistribution over a larger number of members best improved the survival fraction. When the expected resource density was generally insufficient for survival, large groups would split into factions and spread over unoccupied land \citep{antón2012fissionfusion, bird2019huntergatherer, wrangham1993ecological}. Some groups would benefit from upward fluctuations in the resource distribution and lowered competition, allowing for their survival without massive redistribution and resource averaging. In modern populations, it is not always apparent if the resource density in a finite location is sufficient, and mass migration to unoccupied space is not always possible as an act of separation. We model the growth of kinship networks in a spatially-limited environment and discuss the impact of cooperative redistribution on distance-dependent relatedness.

\subsubsection{Simple Diffusion Model Description} 

We first use a resource-independent model to introduce the dynamics and kinship structure resulting from basic local interactions. Agents exist on a finite, toroidal map, $\mathcal{M}_{n \times n}$, of equal grid-side dimension, $n$. Agents experience two life phases – \textit{partner search} and \textit{reproduction}. An initial number of individual agents, $N_{0}$, is distributed at random locations. At each time step, agents migrate in a randomized order. Individual agents will first check their Moore neighborhood for other individual agents. In this case, the agent will move to the occupied neighboring cell, forming a pair. Otherwise, each individual agent or existing pair may move to any empty cell within its Moore neighborhood. Pairs move twice within a single time step.  

At the end of each time step, pairs will reproduce once in random order if there exists an empty neighboring cell to place offspring. The lineage of all agents can be traced through recursive parent referencing. Agents have a preset lifespan, effectively bounding mobility. Pairs die when the first member dies. This is similar to a two-dimensional stepping-stone model with local reproduction and diffusion \citep{kimura1964steppingstone, Dawson90_SteppingStone, Sato83_SteppingStone, BirknerCernyDepperschmidt2015, BirknerGantert2019_Ancestral, BirknerEtAl2024_Coalescence}. 

\subsubsection{Distance-Dependent Relatedness}

In this work, we assume the relatedness between two agents can be determined from their most recent common ancestor (MRCA), $c$. The path sum of the number of generations between the $i$th agent to this common ancestor and that of the $j$th agent to the ancestor is, $k_{c} = g_{i}(c) + g_{j}(c)$. If multiple ancestors exist at this depth, only one is counted. This toy relatedness is expressed as, 
\begin{equation} 
r_{i,j} = \bigg( \dfrac{1}{2} \bigg)^{g_{i}(c) + g_{j}(c) } . 
\end{equation}
It is important to note that in real biological populations, genetic overlap is incredibly high in any differentiated species. We are not describing the evolution of allele distributions, and do not account for multiple common ancestors or pedigree collapse. We are not discussing a formal representation of genetic similarity. Instead, relatedness, $r_{i,j}$, in our model acts as a proxy for a sociological awareness of lineage, which exists in many biological populations. 

For a diffusive random walk through Moore neighborhoods of partial cell occupancy fraction, $\alpha$, the opportunity for motion is decreased. The effective diffusion constant for individual agents, and paired agents, per time step is, 
\begin{equation}
D_{\text{eff}} =
\begin{cases}
\dfrac{3}{8}\,p_{\mathrm{move}}, & \text{individual agent}, \\[8pt]
\dfrac{3}{4}\,p_{\mathrm{move}}, & \text{paired agent}.
\end{cases}
\end{equation}
Here, $p_{\mathrm{move}}$, is the probability of migration to a new cell within the turn, described in Appendix~C. 

The approximate expected relatedness of two agents at a Euclidean distance, $d_E$, is a sum over potential, most-recent generational path distances between two agents, $k$. If reproduction occurs on a timescale, $T_{r}$, the effective timescale separating relatives or agents of a common ancestor is $t_\mathrm{lineage} = kT_{r}$. In the diffusive regime with $\alpha < 0.5$, we might expect from diffusion and spatial coalescence, 
\begin{align}
    \mathbb{E}[r \mid d_{E}] \approx& \sum_{k = 1}^{\infty} \bigg( \dfrac{1}{2} \bigg)^{k} \omega(k;d_{E})P_{\textrm{coal}}(k), \\
    \approx& \sum_{k = 1}^{\infty} \bigg( \dfrac{1}{2} \bigg)^{k}   \frac{d_E}{2 D_{\text{eff}} kT_{r}} \exp\!\left(-\frac{d_E^2}{4 D_{\text{eff}} kT_{r}}\right)    \exp \bigg( -\dfrac{k}{T_{c}} \bigg).
\end{align}
Relatedness at a $k$th generational depth falls as a function of distance, with more diffuse relatives for higher $k$. This reflects the random walks of previous generations, and is represented by the weighting, $\omega(k ; d_{E})$. The contribution at the $k$th generational distance is attenuated by the coalescence of separate ancestral lines. Owing to the finite map size, ancestral lines are likely to merge after a generational time, $\sim$$T_{c}$, lowering the contribution at higher $k$. This is expressed by the probability, $P_{\textrm{coal}}(k)$. This approximation does not reflect anisotropy introduced by walks within Moore neighborhoods or the differing contributions of individuals and pairs. It is intended to illustrate a rough functional form of the steady-state expectation. Details of this analytic description are provided in Appendix~C.

As our population capacity is largely limited by space, spatial jamming plays a substantial role in the dynamics of the system. In the case that most agents are bound in pairs, finite agent lifetime and population turnover leads to $\alpha \approx 0.8-0.9$. This is the subdiffusive or percolative regime, where the expectation of a diffusive walk is not entirely applicable. 

In this work, we primarily compare the results of numerical simulations to understand the relation between distance-dependent relatedness and resource density. As agents have equal mobility along horizontal, vertical and diagonal displacements, we use a Chebyshev metric to describe displacement. 

\section{Results}

\subsection{Resource-Independent Dynamics}

We perform a numerical simulation of our described model. A map of dimension $n = 16$ is used with an initial population of $N_{0} = 256$. As the representation of deep family trees becomes resource intensive, we consider relatedness only out to five generations, $\textrm{max}(k_{c}) = 10$. Agents live for $\Delta T = 10$ time steps. The simulation is evolved for $\Delta T = 1000$ time steps to allow for sufficient mixing relative to the map size. This also provides time for the map to reach population capacity. Agents are then largely limited in reproduction by space. The number of offspring per pair approaches the replacement rate of two. 

\begin{figure}[t]
    \centering
    \includegraphics[width=0.63\linewidth]{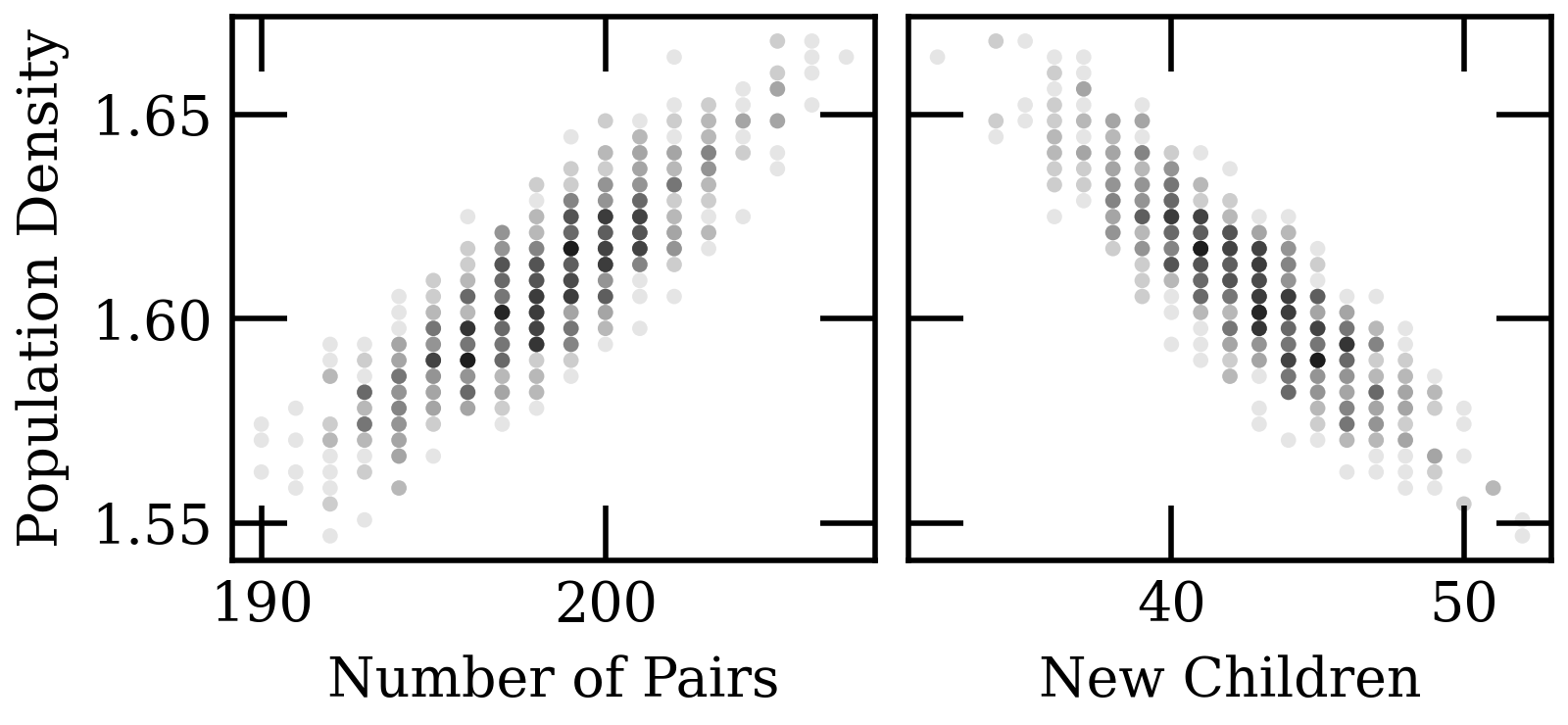}
    \caption{Population density with the number of living pairs and new children per time step. Here, population density represents the total number of agents relative to the number of existing grid spaces. The plots reflect a single simulation with instantaneous statistics from each time step after $\Delta T = 250$. Relative to agent mobility and the small map size, the population has reached an effective steady state after this period. High population densities indicate a large number of pairs and a small number of new children within the time step. This is a spatially-limited population. High density reduces population growth.    }
    \label{fig:simp}
\end{figure}

Variations in $N(t)$ at late times are introduced by inefficient placement, pairing and reproduction. Given a number of grid cells, $K = n \times n$, the grid is maximally occupied when each cell is occupied by a pair of two agents. As an agent and their corresponding pair is removed from the map after a period $T$, there is a continuous population turnover. The number of placed offspring depends on the number of pairs and free cells. Once placed, the probability of pair formation depends on the number and location of other individual agents. We compare the total number of living pairs and number of new children per time step in Figure \ref{fig:simp}.

Assuming a well-mixed population, we can describe an age-structured discrete-time differential equation map, dependent on agent class densities. The total number of agents is a sum over agents and pairs of maximal partner age $a = 0,1, \dots, T-1$,
\begin{equation}
    N(t) = \sum_{a = 0}^{T - 1}N_{\textrm{indiv,a}}(t) + \sum_{a = 0}^{T - 1}N_{\textrm{pair,a}}(t).
\end{equation}

Given the probability of pairing with another individual agent within a Moore neighborhood, the expected number of new agent pairs within the turn can be evaluated, $\mathbb{E}[\Delta N_{\mathrm{pair},a}^{(+)}(t)]$. Similarly, based on the updated map density and the existence of unoccupied spaces, the number of new agents produced within the turn is, $\mathbb{E}[\Delta N_{\mathrm{indiv}}^{(+)}(t)]$. Last, agents are lost through natural death when their age class is dropped. Our discrete time mapping and age class update can then be expressed as, 
\begin{equation}
\left\{
\begin{aligned}
N_{\mathrm{indiv}}(t+1)
&=
\sum_{a=0}^{T-2} N_{\mathrm{indiv,a}}(t)
- 2\,\mathbb{E}\left[\Delta N_{\mathrm{pair}}^{(+)}(t)\right]
+ \mathbb{E}\left[\Delta N_{\mathrm{indiv}}^{(+)}(t)\right],\\[1em]
N_{\mathrm{pair}}(t+1) 
&= 
\sum_{a=0}^{T-2} N_{\mathrm{pair,a}}(t) 
+ \sum_{a=0}^{T-2} \mathbb{E}\left[\Delta N_{\mathrm{pair},a}^{(+)}(t)\right].
\end{aligned}
\right.
\end{equation}
\begin{equation}
\begin{aligned}
N_{\mathrm{indiv,}0}(t+1) &= \mathbb{E}\left[\Delta N_{\mathrm{indiv}}^{(+)}(t)\right], \\
N_{\mathrm{indiv,}a+1}(t+1) &= N_{\mathrm{indiv,}a}^{\star}(t),
\\
N_{\mathrm{pair,}a+1}(t+1) &= N_{\mathrm{pair,}a}^{\star}(t),
\qquad a=0,\dots,T-2.
\end{aligned}
\end{equation}
Here, $N^{*}(t)$, represents the post-interaction population, prior to aging. Ages are then incremented, forming classes for the next time step. 

In general, the steady-state population is spatially-limited. We provide an in-depth discussion in Appendix~D, where both the discrete-time mapping and mean-field ordinary differential equation (ODE) system are described.

\subsection{Resource-Dependent Dynamics}

We now assume that agent survival is additionally tied to the random distribution of resources. At each time step, after potentially moving to a new grid cell, an integer number of resources is drawn from a Poisson distribution of mean $\mu$ for each agent. Agents must have at least $\phi = 1$ resources for survival within the turn. Agents with insufficient resources are removed from the map at the end of this pass. Agents may then reproduce, utilizing the present free space. Agents or pairs of greatest age, $T$, are removed at the end of the turn. Last, any residual resources are removed from each agent.

\begin{figure}
    \centering
    \begin{subfigure}{\textwidth}
        \centering
        \includegraphics[width=0.6\linewidth]{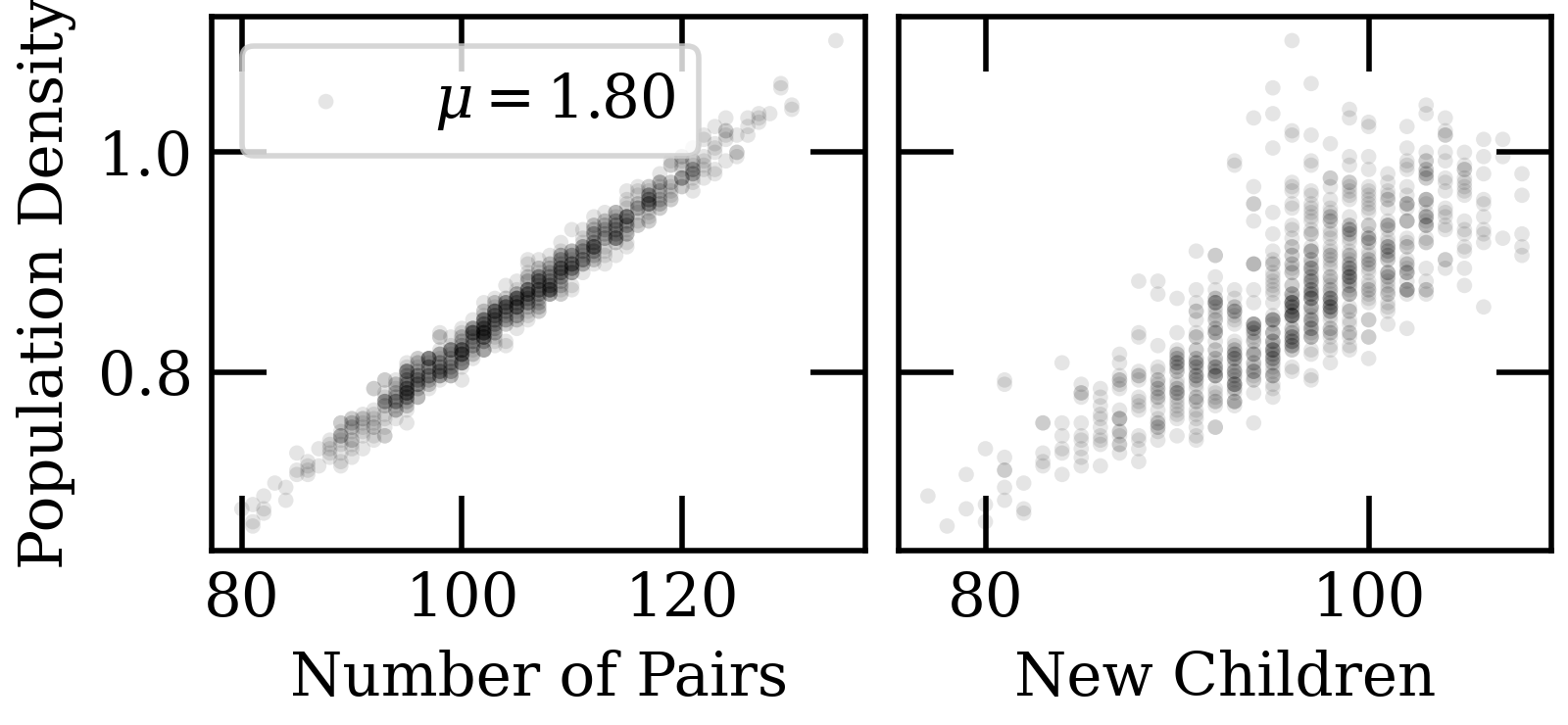}
    \end{subfigure}

    \begin{subfigure}{\textwidth}
        \centering
        \includegraphics[width=0.6\linewidth]{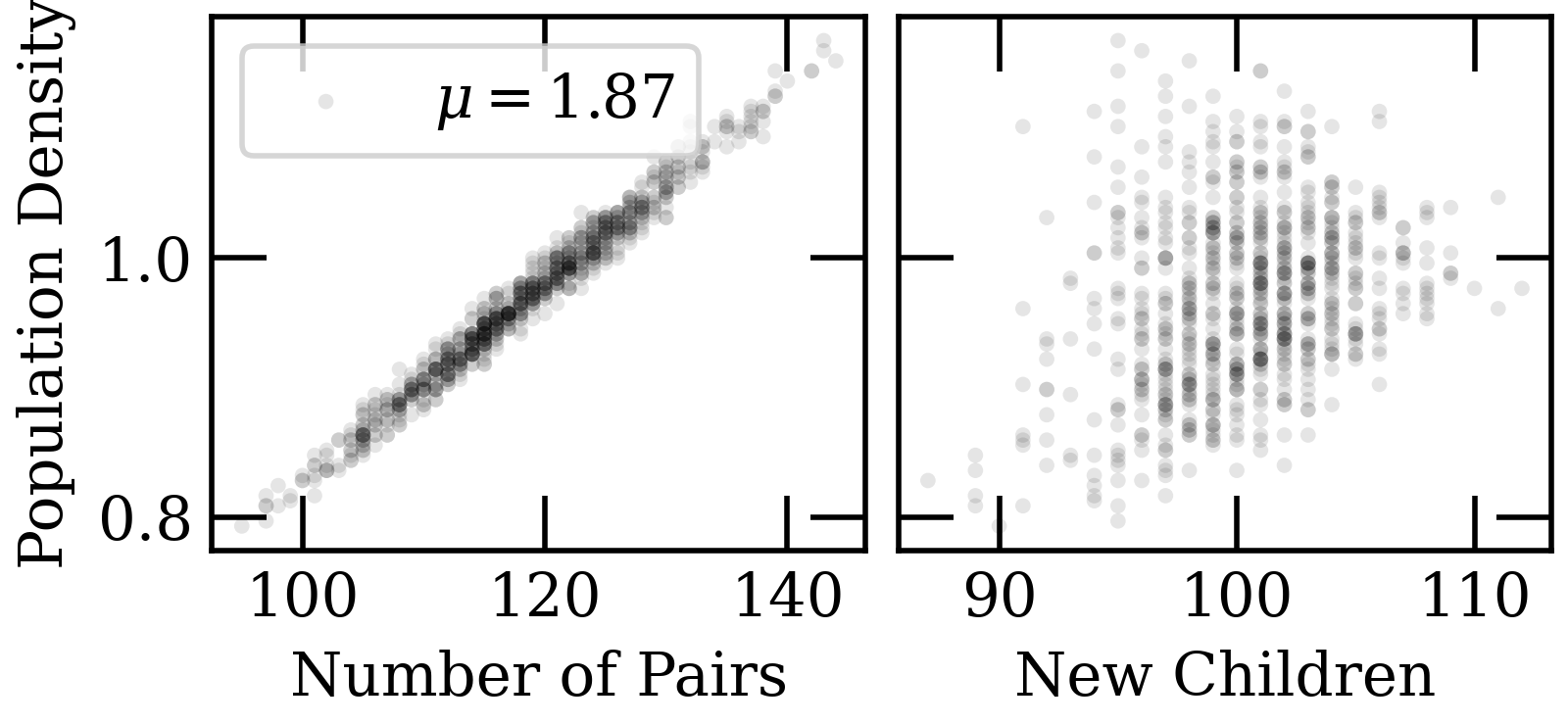}
    \end{subfigure}

    \begin{subfigure}{\textwidth}
        \centering
        \includegraphics[width=0.6\linewidth]{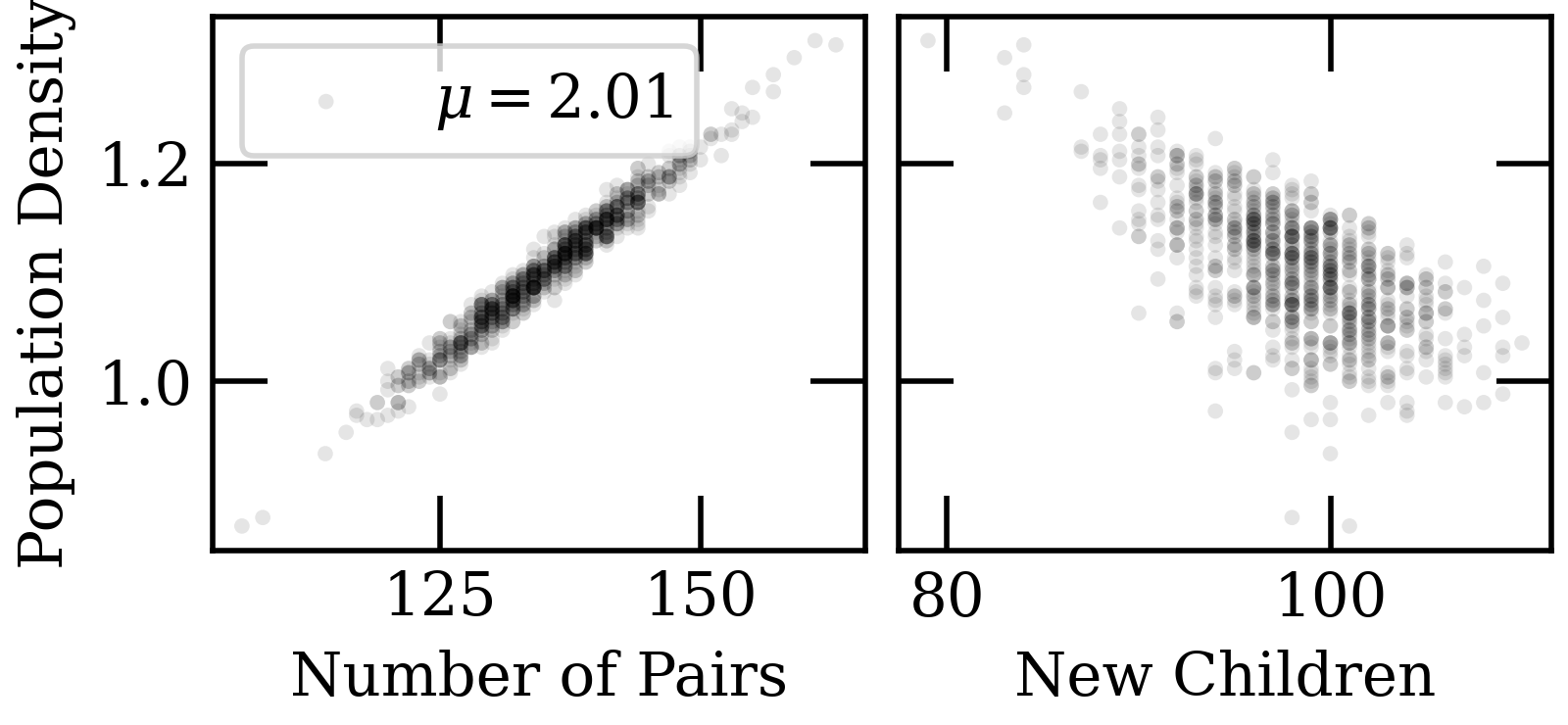}
    \end{subfigure}

    \caption{  Resource-dependent population density with the number of post-redistribution living pairs and new children per time step. Here, population density represents the total number of agents relative to the number of existing grid spaces. Each plot row reflects a single simulation with instantaneous statistics from each time step after $\Delta T = 250$. The resource mean of $\mu \approx 1.8$ was determined numerically as a critical point for the system, separating extinction from population persistence. Prior to this critical point, a lower, non-zero population density may exist for an extended period before extinction through fluctuations. Just beyond the critical point, this population is mortality limited, and a higher number of pairs leads to a higher number of children. At higher $\mu$ values, the population is largely impacted by spatial saturation, and a larger number of pairs leads to a lower number of children.  }
    \label{fig:bif}
\end{figure}

The chosen resource update scheme makes population survival dependent on the instantaneous resource distribution of that turn. Resources do not build with time and are not stored. Additionally, agent updates are randomized by specific agent number, but ordered by the type of operation. This most closely resembles seasonal variations, where variable resources are collected cyclically and later redistributed after deliberation. This choice decouples resource growth and population growth.

The initial probability of survival is, 
\begin{equation}
    P_{\textrm{surv}}(\mu, \phi)|_{\phi = 1} = 1 - e^{-\mu} .
\end{equation}
The agent and pair discrete-time mapping becomes, 
\begin{align}
N_{\mathrm{indiv}}(t+1)
&=
P_{\mathrm{surv}}(\mu,\phi)
\left[
\sum_{a=0}^{T-2} N_{\mathrm{indiv},a}(t)
\right]
\nonumber\\
&\quad
- 2\,\mathbb{E}\!\left[\Delta N_{\mathrm{pair}}^{(+)}(t)\right]
+ P_{\mathrm{surv}}(\mu,\phi)^{2}
\,\mathbb{E}\!\left[\Delta N_{\mathrm{indiv}}^{(+)}(t)\right],
\\[1em]
N_{\mathrm{pair}}(t+1)
&=
P_{\mathrm{surv}}(\mu,\phi)^{2}
\left[
\sum_{a=0}^{T-2} N_{\mathrm{pair},a}(t)
+
\sum_{a=0}^{T-2}
\mathbb{E}\!\left[\Delta N_{\mathrm{pair},a}^{(+)}(t)\right]
\right].
\end{align}
The expectation for the number of added individuals also depends on the survival of both paired agents, gaining a dependence $\propto P_{\mathrm{surv}}(\mu,\phi)^{2}$ relative to the resource-independent case. The availability of free spaces for children also now reflects the reduced population density after resource death. We further describe this mapping as well as the complimentary mean-field ODE within Appendix~E. 

Population size exhibits a critical threshold near $\mu \approx 1.8$, corresponding to $P_{\textrm{critical}}$. Below this critical value, extinction is the only stable equilibrium. Beyond $P_{\textrm{critical}}$, a persistent, stable population becomes possible. Near $P_{\textrm{critical}}$, within our finite, discrete-time numerical simulation, stochastic fluctuations may allow for a transient, low-density population before extinction. We provide bifurcation-style plots for relevant quantities for $P_{\mathrm{surv}} \sim P_{\textrm{critical}}$ and $P_{\mathrm{surv}} > P_{\textrm{critical}}$ in Figure \ref{fig:bif}. As $\mu$ increases, the population transitions towards spatial saturation.

\subsection{Effect of Redistribution on Expected Relatedness, Kinship Structure }

Cooperation or wealth redistribution is commonly expected through kinship. Ecological systems show preferential or ranked cooperation based on both relatedness, $r$, as well as distance, $d$. The level of localization of such dependence should be viewed generally and may vary based on system or species. Importantly, the existence of any localization or preference within a distributed system alters the dynamics and can lead to characteristic spatial structure. 

We consider that the distance, $d$ and relatedness $r$ between agents plays a role in the opportunity for resource transfer. Similar to a flux over a toroidal plane, we suggest that the fraction of excess resource transfer between $i$th and $j$th agents of MRCA, $c$,

\begin{equation}
\rho_{i,j} =
\begin{cases}
A, 
& \text{if the \(i\)th and \(j\)th agents are paired}, \\[8pt]
A  \ \left( \dfrac{1}{2} \right)^{g_i(c) + g_j(c)}
\left( \dfrac{1}{d_{i,j}} \right),
& \text{otherwise}.
\end{cases}
\end{equation}
The fraction of excess resources transferred decreases both with relatedness and planar flux. A specific value of $\rho_{i,j}$ can be viewed as an agent-to-agent edge weight within our kinship network. $A$ is a constant, effectively controlling the size of population relied upon for excess resource transfer. Agents with highest transfer fraction are asked first. 

While this general protocol for agent-to-agent redistribution could be adaptable or partially learned, there is generally an intrinsic preferential relationship between parent and children within any reproducing population, or between a localized and time-dependent group of caretakers and dependents. We focus on this asymmetry and explore the implications of this static dependency structure. 

After all agents receive their initial resources, if an agent does not have the desired number, they will make requests to other related members of the population until their need is satisfied. Agents complete this request process in a randomized order. As resources are refreshed with each time step, we are able to focus on the structures driven through redistribution-dependent survival. 

We perform a suite of simulations to characterize the system and resulting network structures. As population survival and growth depends on an existing network structure, we begin each simulation by developing a resource-independent population for 1000 time steps. After this period, we include resource-dependence and redistribution, and track the population for another 1000 time steps. 

\begin{figure}
  \centering
   \begin{subfigure}{\textwidth}
    \centering
    \includegraphics[width=0.75\linewidth]{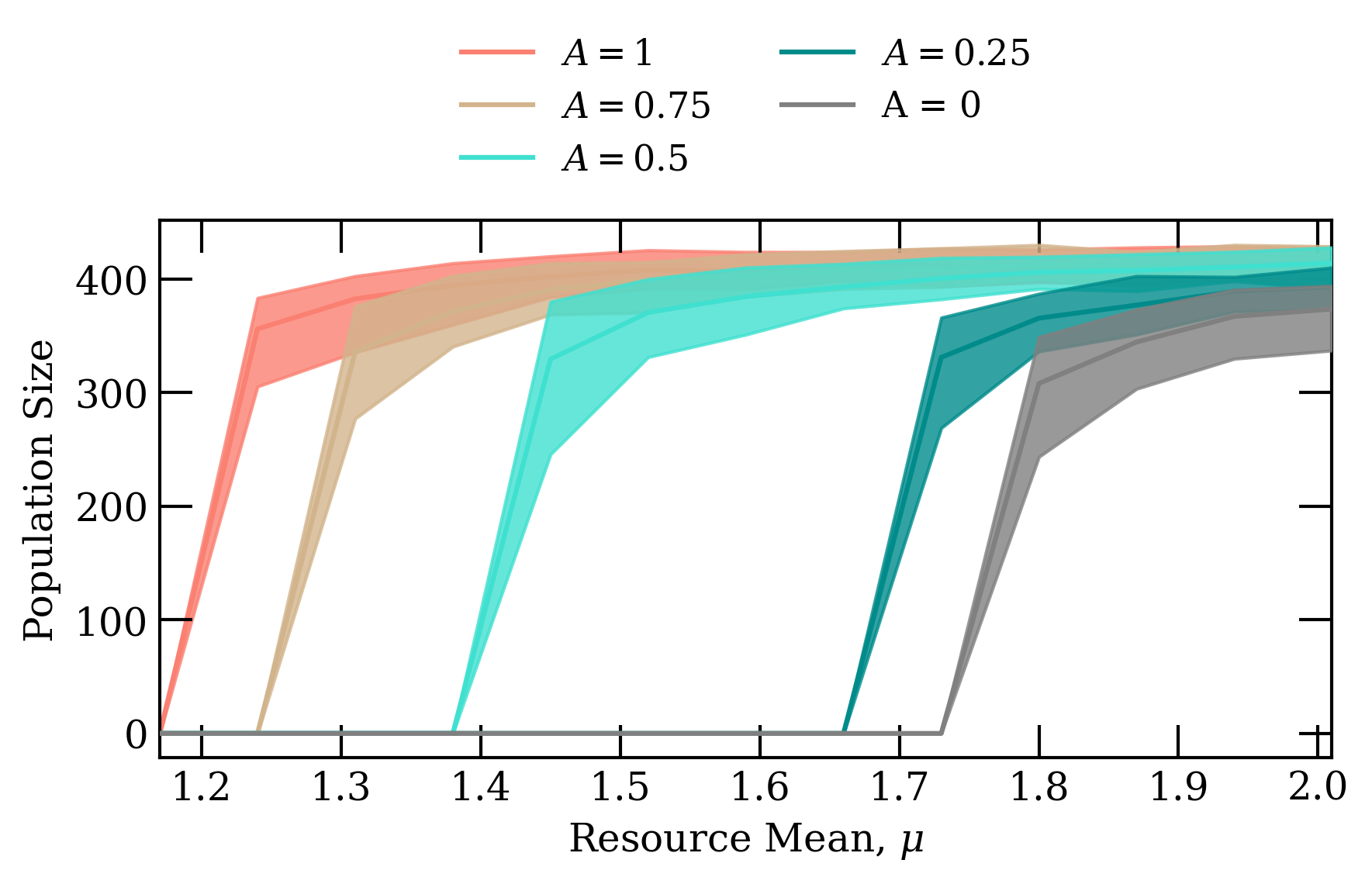}
  \end{subfigure}

  \vspace{0.5cm}

  \begin{subfigure}{\textwidth}
    \centering
    \includegraphics[width=0.75\linewidth]{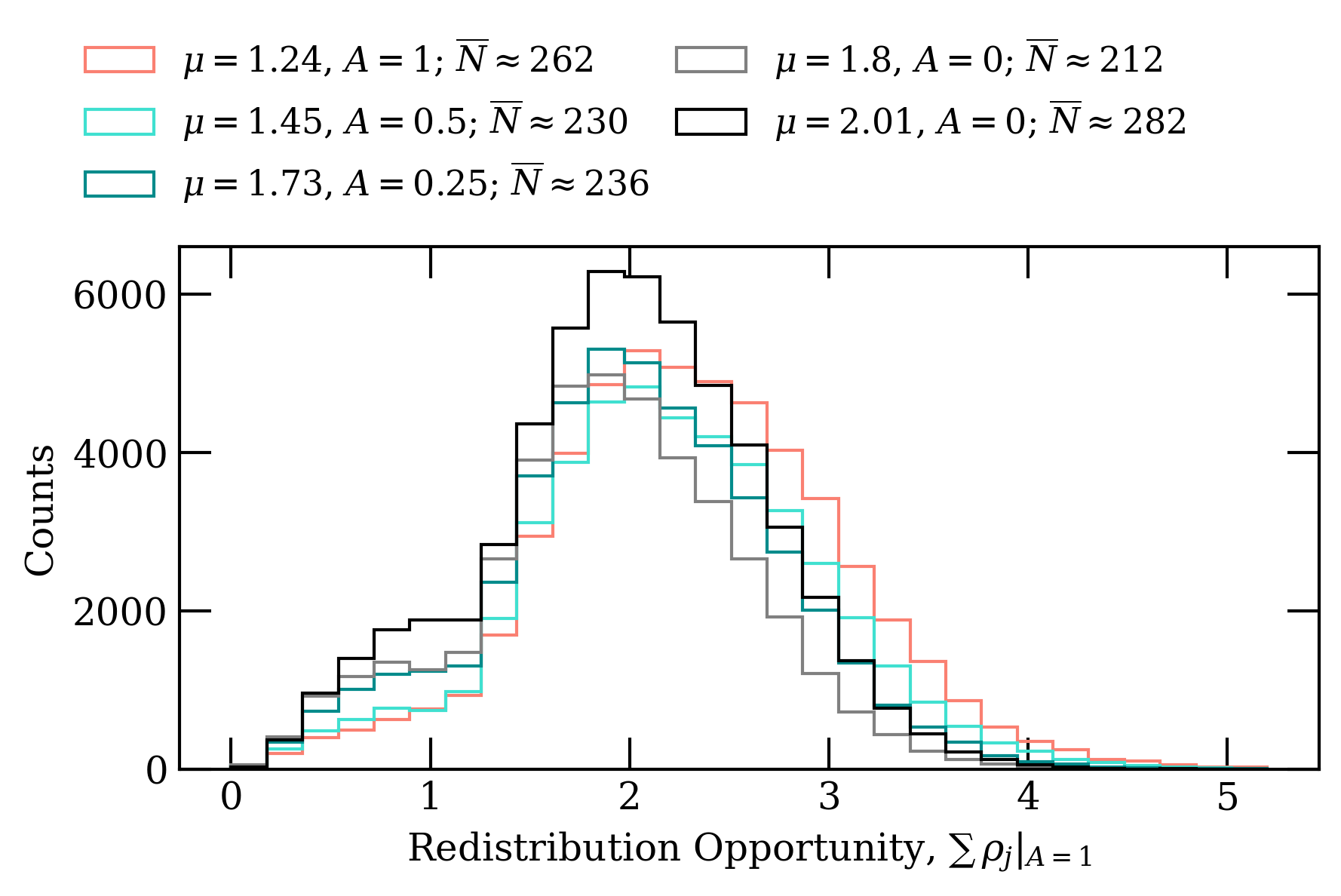}
  \end{subfigure}
  \caption{Population size and agent redistribution opportunity for several values of $A$. The case of $A = 0$ represents a resource-dependent population without redistribution. On the top, we provide population size as a function of $\mu$ at late time, $\Delta T = 2000$. With each resource value and redistribution scheme, we plot the median and $98 \%$ central containment region of an ensemble of 50 simulations. The transition to a persistent population is rapid, and the course resource-mean sampling only serves to give a rough localization of $P_{\textrm{critical}}$. Once a persistent, stable population is possible, the median population size increases with $\mu$ with the transition towards spatial saturation. On the bottom, we provide several distributions of summed agent redistribution opportunity, $ \sum_{j} \rho_{j}|_{A = 1}$. The labeled value of $A$ within the legend represents the value used in simulation, while $A = 1$ is evaluated here for clearer statistical comparison. The resulting distributions reflect only agent generational and spatial familial proximity. These are aggregate distributions from ensembles of 200 simulations. We find that these distributions are lifted towards higher values when redistribution is possible, suggesting more organized or clustered populations.  }
  \label{fig:full_pop_surv}
\end{figure}

The impact of redistribution is to lower $P_{\textrm{critical}}$ and $\mu$ required for persistent populations. We provide a comparison of population survival as a function of $\mu$ for several values of $A$ and the case without resource redistribution in Figure \ref{fig:full_pop_surv}. Further, agents are now dependent on their node strength or local network structure for survival. Similar to our simplified hub-and-spoke model, highly connected agents draw on the fringe population (effective spokes), and are more likely to survive and reproduce. These fringe members essentially act as resource donors, supporting family members with demand, but receiving disproportionate total support. Fringe members have shorter average lifetimes, subject to resource volatility, while hubs are able to survive by redistribution for longer periods. Such differences in network structure occur naturally due to its stochasticity. 

Agents that successfully reproduce pass on their network structure, potentially transmitting this preferential survival within the population for future generations. As the opportunity for agent reproduction and increased lineage size depends on the existing familial size and proximity, we might expect an approximate power-law or scale-free relation for the distribution of developed familial sizes. Similarly, agents which are closest to these effective hubs should have a higher opportunity for redistributive requests to other agents, which can be represented by a sum over each $j$th related agent, $\sum_{j} \rho_{j}$. The highest summed opportunity values experienced by these selected agents should be higher than those realized in comparable simulations without redistribution, due to this structural growth feedback. 

As a demonstration of structural differences between populations developed with and without cooperative redistribution, we provide the aggregate distribution of redistribution opportunity, $\sum_{j} \rho_{j}$, for several values of $A$, and the case of no redistribution from an ensemble of simulations. As this value could also depend on the average number of living population members at a given time, $\overline{N}$, we select the sampled resource mean, $\mu$, closest to $P_{\textrm{critical}}$ in each case. These distributions are presented in Figure \ref{fig:full_pop_surv}. With smaller populations, the mean family size and opportunity for redistribution could be slightly smaller, generally leading to a smaller $\sum_{j} \rho_{j}$ reached for the population. To demonstrate that this potential effect does not greatly impact our comparison of distributions, in the case without redistribution, we have provided both $\mu = 1.8$ and $\mu = 2.01$, where the latter resource mean corresponds to a substantially higher mean population, with a similar distribution shape as $\mu = 1.8$. We also note that extended relatives of generational distance beyond $\textrm{max}(k_{c}) = 10$ should exist at large spatial distances, and are not expected to substantially impact $\sum_{j} \rho_{j}$ or the qualitative comparison of distributions. 

We find that distributions representative of redistribution correspond to longer tails for high values of $\sum_{j} \rho_{j}$. The distribution appears less peaked, reflecting increased network node strength heterogeneity. Additionally, the weight or mean of these distributions tends towards higher values of redistribution opportunity, suggesting an overall greater level of kinship organization compared to the case without redistribution. These characteristics appear more strongly for populations heavily dependent on redistribution. A smaller scale factor, $A$, corresponds to a higher value of $P_{\textrm{critical}}$, where redistribution is less relied upon. We also point out that these results reflect changes in the extended kinship network as opposed to potential differences in pairing. This is further demonstrated in Appendix~F. 

\begin{figure}
    \begin{subfigure}{\textwidth}
        \centering
        \includegraphics[width=1.0\linewidth]{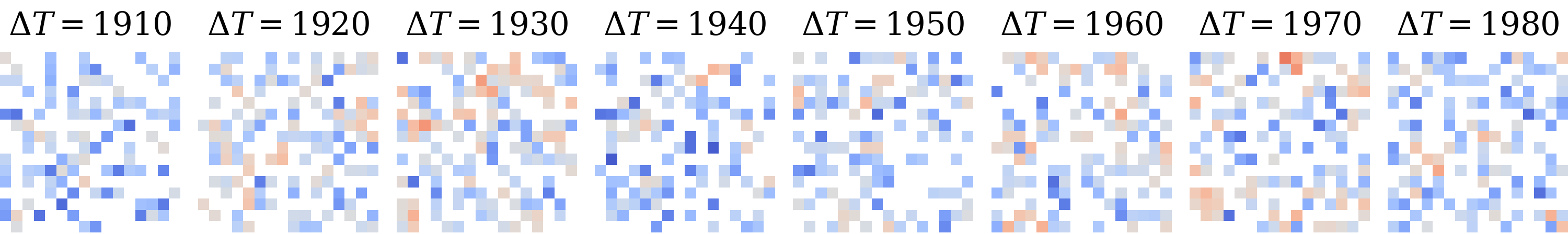}
    \end{subfigure}
    \\[0.25cm]
    \begin{subfigure}{\textwidth}
        \centering
        \includegraphics[width=1.0\linewidth]{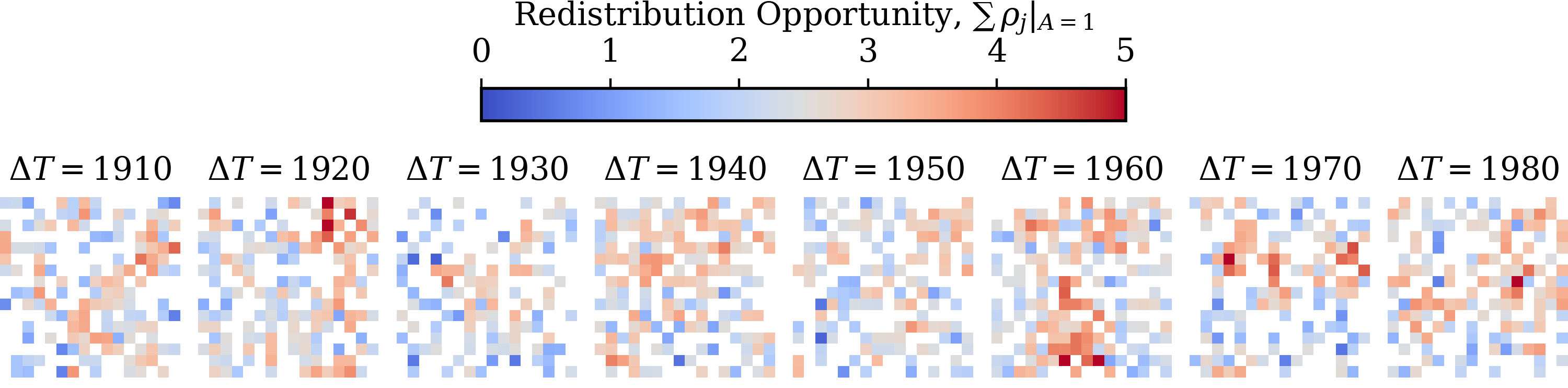}
    \end{subfigure}
    \caption{Example spatiotemporal evolution of agent redistribution opportunity, $ \sum_{j} \rho_{j}$, with and without redistribution. We plot agent redistribution opportunity, $ \sum_{j} \rho_{j}|_{A = 1}$ as a function of grid location for several time stamps at late simulation times. If a pair is present, the greater redistribution opportunity is plotted. In the top row, we plot an example of a resource-dependent population without redistribution with $\mu = 1.8$. At the bottom, we plot a population with $\mu = 1.24$ and $A = 1$. The colorbar is shared between the two rows, and values are clipped to a maximum of 5. The population with redistribution exhibits clusters of high $ \sum_{j} \rho_{j}$, suggesting the development of tighter family structures.   }
    \label{fig:cluster}
\end{figure}

In Figure \ref{fig:cluster}, we provide an example of spatiotemporal evolution both for the case with and without redistribution as a function of summed redistribution opportunity. We find some visible clustering and pattern formation. Increased clustering of high redistribution-opportunity agents is notable relative to potential spatial correlations observed in the case without redistribution. These structures dissipate or migrate quickly. They also appear denser, suggesting recent familial growth. Reproduction for agents within these local structures may become spatially-limited. This provides a braking mechanism for the rapid growth of individual families. Instead, close family that is nearby, well-connected, but in lower-density zones may reproduce. This ensures migration of the hub or over-density.

\begin{figure}
  \centering
  \includegraphics[width=1.0\textwidth]{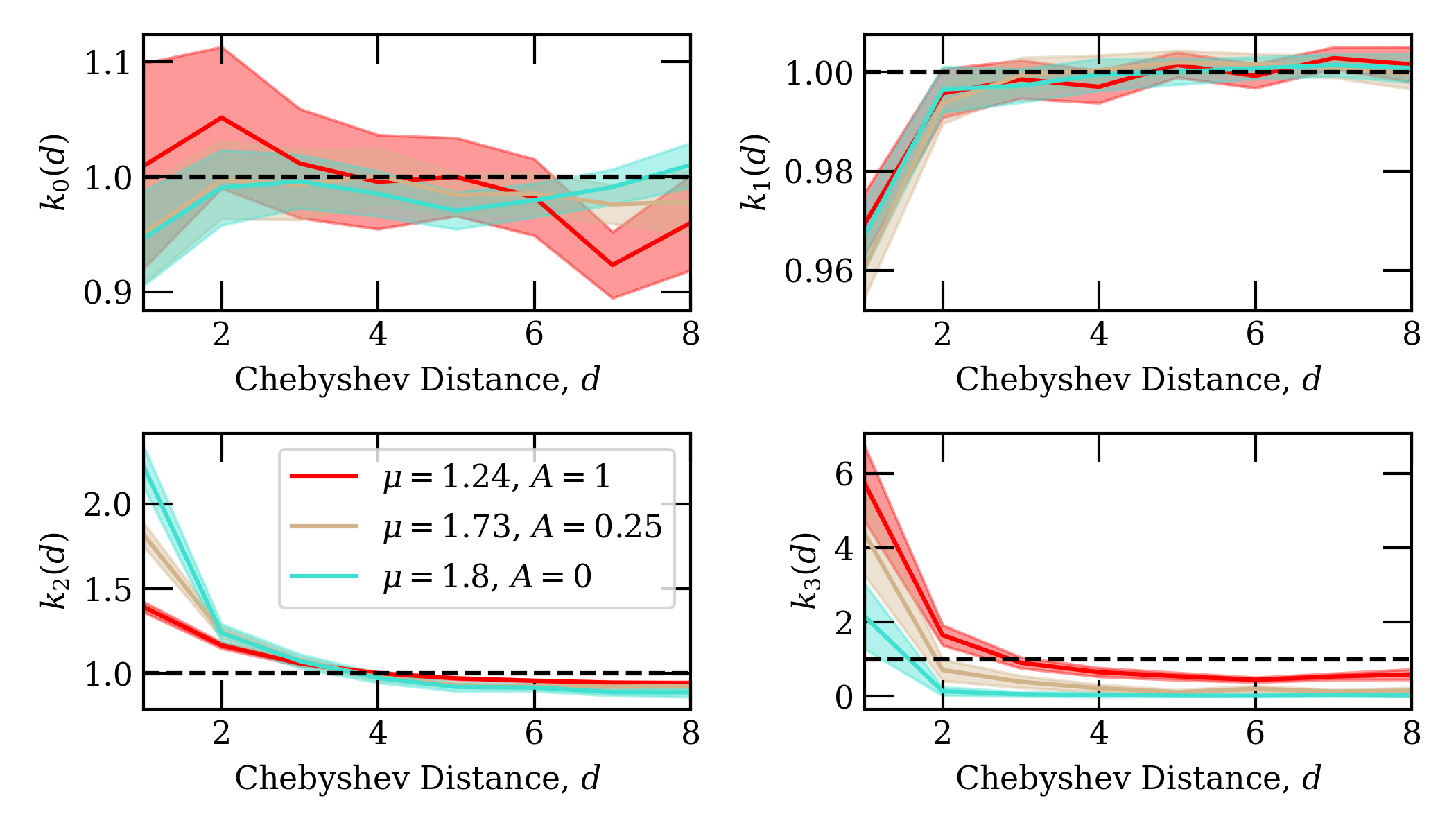}
  \caption{Marked two-point correlation as a function of Chebyshev distance with and without redistribution. We perform a marked two-point correlation of summed agent redistribution opportunity $\sum_{j} \rho_{j} |_{A = 1}$ values for four equally spaced bins in $[0,5]$. While the simulation is performed with a specific value of $A$, we set this scale factor, $A = 1$, in plotting and analysis for a direct comparison of spatial-generational properties of the population. In each redistribution case, we consider an ensemble of 300 simulations at late time, $\Delta T = 2000$. We plot the mean two-point correlation ratio, $\overline{k}_{b}(d)$ for each bin, $b$, with a solid line. The 95$\%$ confidence interval on the mean is indicated by the shaded regions. Populations with redistribution have substantial clustering at highest $\sum_{j} \rho_{j} |_{A = 1}$ values.  }
  \label{fig:corr}
\end{figure}

Last, we also compare structural clustering with a marked two-point correlation analysis of agent $\sum_{j} \rho_{j}$ values. The assumed values of $A = 1$ and $A = 0.25$ are evaluated as well as the case without redistribution. With each redistribution scheme, we consider the simulation state at $\Delta T = 2000$ for an ensemble of 300 simulations. For a direct comparison, we set the scale factor, $A = 1$, for a generational and spatial-distance dependent analysis. 

We consider the value $\sum_{j} \rho_{j} | _{A = 1}$ over the range $[0,5]$, which we discretize into four evenly spaced bins. With each bin, $b$, we determine the number of agent value-pairs $DD_{b}(d)$ with values lying in the bin, $b$, and separated by Chebyshev distance, $d$. Pair partners within the same grid cell are not considered. 

We define a null model as the expected number of value-pairs from values, $\sum_{j} \rho_{j} | _{A = 1}$, randomly permuted across agents. Given the spatial distribution realized by each simulation, we randomized agent values 100 times to determine the count expectation per bin $b$, $RR_{b}(d)$. 

The marked two-point correlation ratio of bin $b$ is then, 
\begin{equation}
    k_{b}(d) = \dfrac{DD_{b}(d)}{RR_{b}(d)}.
\end{equation}
Given this definition, $k_{b} > 1$, reflects an excess of value-pairs at distance $d$ relative to a randomized realization. $k_{b} < 1$ reflects a suppressed number of value-pairs. As the toroidal grid has periodic boundaries, additional corrections for map geometry are not required. 

In general, as the locations of agents are not permuted, the underlying density produced by the reproduction dynamics is still represented. This test instead provides a measure of the level of clustering based on $\sum_{j} \rho_{j} | _{A = 1}$ alone. The results are presented in Figure \ref{fig:corr}. 

At lowest bin values, we find essentially no spatial correlation between agents, indicating these agents are diffuse throughout the map. However, the confidence intervals are broad due to the lower sample statistics relative to other bins. In the second bin, $k_{1}(d)$ shows noticeable suppressed clustering at short distances. At large distances, poorly-related agents are diffuse, while at short distances, poorly-related agents are unlikely to cluster. In the third bin, $k_{2}(d)$, these moderate $\sum_{j} \rho_{j} | _{A = 1}$ values are generally correlated at short distances, especially in the case without redistribution. This indicates the amount of spatial-generational clustering naturally reached through our reproduction dynamics. However, redistribution-capable populations have lower relative clustering at these values. Instead, we see in the highest value bin, $k_{3}(d)$, that the greatest clustering at short distances is observed in the case of redistribution. These well-related agents are most clustered, potentially with more diffuse, less-related agents around them.

\section{Discussion}

In this work, we have developed a process for threshold resource redistribution across agents within a network. We have presented the mean field expectation and finite population-size corrections for a fully connected network, as well as numerical simulations. Further, we have discussed such a process over a hub-and-spoke network. Owing to the thresholding of resource redistribution, a network topology of varied node strength leads to preferential node survival. Highly connected agents or hubs have a larger opportunity to draw on the excess of connected spokes, while spokes draw collectively on the potential excess of the hub. 

We have then explored the impact of cooperation or resource redistribution within spatially-limited kinship networks. We first presented a simple model for the growth of kinship networks on a finite, toroidal grid. The diffusion of walking agents and finite map results in distance-dependent relatedness as an approximate sum of spatial Gaussians with generational intensity attenuated by the timescale of ancestral coalescence. Next, we introduced the ability for resource redistribution with excess resource transfer fraction, $\rho$, dependent both on the distance and relatedness between agents. We found that populations may persist for lower resource densities at the cost of overall increased agent node strength and node strength heterogeneity. A greater dependence on redistribution leads to the preferential survival of highly-connected agents (hubs), those with large, spatially and ancestrally close families. Poorly connected or fringe family members (spokes) still exist and provide a source of potential resource excess for these hubs. Last, we performed a two-point correlation statistical test, which demonstrated spatial-clustering at high $\sum_{j} \rho_{j}$ and low distances, with anti-correlation at large distances. While a formal analysis of network centrality was not performed, the high-tailed agent node strength distribution represented by $\sum_{j} \rho_{j}$ and spatial correlations are suggestive of a more centralized kinship structure or the emergence of locally centralized clusters. 

In general, an agent's opportunity for redistribution, survival and reproduction, a reflection of the strength of kinship connections, is more likely to grow if it is already larger relative to the  population. This feedback drives the growth of structure within our model. This scheme supports our heavy-tailed distribution of agent node strengths and the interpretation of increased centrality. 

In our simulation, spatial saturation caps the growth of individual agent kinship networks. While not explicitly considered in this work, representation of the initial population through lineage at late times may still approximate a power law, even though growth on the span of individual agent lifetimes is limited. 

The main result of this work is a demonstration of the relation between resource density, cooperation and kinship structure. Resource redistribution allows for persistent populations at lower expected resource means, at the cost of apparent, increased kinship network centralization. The emergence of clustering for agents with the greatest node strength was robust across a range of transfer scale factors, and appeared strongest when a greater dependence on redistribution was required. 

The relevance of this result largely depends on the realism of our redistribution model. A number of choices represent model systematics in this work –– including the ability to accept resources only up until a desired energy threshold, fractional excess sharing as opposed to a complete transfer of a needed resource excess, and the ordering of transfer requests. Additionally, the exact function of kinship and spatial closeness may vary in real applications. Still, our toy model for redistribution provides an illustration of the basic dynamics and structure possible and likely in many real biological and sociological networks. 

We have assumed a static form for the transfer fraction, $\rho$, within this work, though it is less clear how such behaviors evolve within natural ecological populations. It seems feasible to identify some preferences for redistribution with biological origins or spatiotemporal limitations and other preferences with a cultural or learned origin. It is difficult to map the transition between these influences. We save these questions for future work. 

It is worth considering how the size and timescale of kinship clusters and over-densities depends on both a generalized initial resource distribution and a generalized redistribution scheme. Allowing the retention of resources by agents or locally within the map may also be appropriate to describe resource-dependent structures on more natural timescales, as well as agent-resource population coupling. 

We have introduced the relation between locally centralized kinship clusters and reproduction. This could be interpreted as an Allee-like mechanism, where network density promotes growth. We have not yet characterized the potential impacts of these transient structures on the system dynamics. 

This partially stylized simulation is an example of an adaptive network. Node or agent state impacts network structure, while the network topology itself shapes dynamics on the nodes. There is a feedback process between agent network structure, survival, and the growth of new connections. While our results are obtained numerically, analytical descriptions would be valuable, and would provide insight into limiting cases of resource redistribution over biological and ecological networks, and within related sociological systems. Specifically, a model of coupled node and edge dynamics for general resource-flow protocols in spatially-constrained, reproducing populations could characterize the resulting clustering and asymmetries in transport. We leave this extension for future work. 

\appendix

\section{Fully Connected, Finite Size Corrections}

In the case of a finite population size, $N$, the variance, $\mu$, of our Poisson distribution contributes an overall variance to the total supply and demand, now relevant along with the mean expectation. Our total demand, $D_{N}$ and supply $S_{N}$ are such that, 
\begin{align}
    D_{N} &= \sum_{i = 1}^{N} d_{i}, \\
    S_{N} &= \rho\sum_{i = 1}^{N} s_{i}. \\
\end{align}
The expected means are, 
\begin{align}
    \mathbb{E}[D_{N}] &= ND = N  \mathbb{E}[(\phi - r)_{+}], \\
    \mathbb{E}[S_{N}] &= NS = N \rho \mathbb{E}[(r - \phi)_{+}].
\end{align}
The total supply and demand variances add from individual agents, 
\begin{align}
    \textrm{Var}(D_{N}) &= N \sigma_{D}^{2} = N \textrm{ Var}( (\phi - r)_{+} ), \\
    \textrm{Var}(S_{N}) &= N \sigma_{S}^{2} = N \rho^{2} \textrm{ Var}( (r - \phi)_{+} ).
\end{align}
The condition for full survival, $X_{N} = S_{N} - D_{N} \geq 0$. Then, by the central limit theorem \citep{kallenberg2002foundations}, as $N$ becomes large, scaled fluctuations of $X_{N}$ around the mean field value follow, 
\begin{equation}
    \dfrac{X_{N} - N(S - D) }{ \sqrt{N(\sigma_{S}^{2} + \sigma_{D}^{2})}} \longrightarrow \mathcal{N}(0,1). 
\end{equation}
Here, $\mathcal{N}(0,1)$ is the standard Gaussian distribution with mean of zero and variance of one. The probability of full survival is then, 
\begin{equation}
    \mathcal{P}(N) =  \mathbb{P} (X_{N} \geq 0) \approx \Phi \left( \dfrac{\sqrt{N}(S - D) }{ \sqrt{\sigma_{S}^{2} + \sigma_{D}^{2}} } \right).
\end{equation}
$\Phi$ is the cumulative density function (CDF) of our Gaussian distribution. In the case of finite $N$, the probability of full survival is instead a smoothed sigmoid function centered around $X_{N} = 0$, similar to finite-size effects in critical phenomena \citep{binney1992theory}. The transition width is of scale $\sim$$1/\sqrt{N}$, corresponding to a step function as $N \rightarrow \infty$.

We can also define a fraction of surviving deficit agents in the case of insufficient resources, 
\begin{equation}
    f_{N} = \dfrac{ \textrm{ number of surviving agents}}{N} = \textrm{min} \bigg( 1, \dfrac{S_{N}}{D_{N}} \bigg).
\end{equation}
In the partial survival regime (far from criticality and redistribution truncation), if we describe the demand and supply in terms of fluctuations around their means, we have $S_{N} = NS + \delta S_{N}$, $D_{N} = ND + \delta D_{N}$. Taking the expectation, and expanding around the mean field, 
\begin{equation}
    \mathbb{E}[f_{N}] =  \mathbb{E}\bigg[ \bigg(  \dfrac{S}{D} \bigg( 1 + \dfrac{\delta S_{N}}{NS} -  \dfrac{\delta D_{N}}{ND} \bigg) + O(N^{-1})  \bigg) \bigg].
\end{equation}
As the expectation for both supply and demand fluctuations is zero, we consider the quadratic term, 
\begin{align}
    \mathbb{E} \bigg[  \dfrac{ (\delta D_{N})^{2} }{N^{2} D^{2}} \bigg] &= \dfrac{\sigma_{D}^{2}}{ND^{2}}, \\
    \mathbb{E} \bigg[  \dfrac{ (\delta S_{N})^{2} }{N^{2} S^{2}} \bigg] &= \dfrac{\sigma_{S}^{2}}{NS^{2}}.
\end{align}
The expected survival fraction is then, 
\begin{equation}
    \mathbb{E}[f_{N}] = \dfrac{S}{D} - \dfrac{1}{2N} \bigg(  \dfrac{\sigma_{S}^{2}}{S^{2}} +  \dfrac{\sigma_{D}^{2}}{D^{2}} \bigg) + O(N^{-3/2}).
\end{equation}

\section{Heterogeneous, Finite Size Corrections}

If we consider a smaller population of size $N$, variance in the total expected spoke supply and demand becomes important. Again, the variance of the resource distribution expected for all agents is $\mu$. If a single spoke or hub has demand $d_{i} = (\phi - r_{i})_{+}$, we define $\mathbb{E}[d] = D_{*}$ and $\textrm{Var(d)} = \sigma^{2}_{D}$. Following the central limit theorem, the total node demand, 
\begin{equation}
    D^{\textrm{tot}}_{s} \approx (N - 1)D_{*} + \sqrt{N - 1} \textrm{ } \sigma_{D} \textrm{ }\eta_D.
\end{equation}
Here, $\eta_D \sim \mathcal{N}(0,1)$. Similarly, for a node supply of $s_{i} = (r_{i} - \phi )_{+}$, we have  $\mathbb{E}[s] = S_{*}$ and $\textrm{Var(s)} = \sigma^{2}_{S}$. The total spoke supply is then, 
\begin{equation}
    S^{\textrm{tot}}_{s} \approx (N - 1)S_{*} + \sqrt{N - 1} \textrm{ } \sigma_{S} \textrm{ }\eta_S.
\end{equation}
Again, $\eta_S \sim \mathcal{N}(0,1)$.

As supply and demand variance is associated with both the hub and spokes, hub survival is possible if,
\begin{align}
    (N - 1)S_{*}& - D_{*} + \sqrt{ (N - 1)\sigma_{S}^{2} + \sigma_{D}^{2}} \eta \geq 0, \\ 
    \eta &\geq \dfrac{D_{*} - (N - 1)S_{*}   }{ \sqrt{ (N - 1)\sigma_{S}^{2} + \sigma_{D}^{2}} }.
\end{align}
$\eta \sim \mathcal{N}(0,1)$. The probability of hub survival is then, 
\begin{equation}
    \mathcal{P }_{h}  = \Phi \left( \dfrac{ (N - 1)S_{*}  - D_{*}}{ \sqrt{ (N - 1)\sigma_{S}^{2} + \sigma_{D}^{2}} } \right).
\end{equation}

\section{Distance-Dependent Relatedness}

If agents are provided with sufficient resources, mobility is bounded only by the finite agent lifetime and the occupancy of neighboring grid cells. In the sense that agents may reproduce, these random walks are continued through lines of lineage. Given a grid occupancy, $\alpha$, finite agent lifetime, $T$, and mean reproduction time, $T_{r}$, we can consider an analytical description of the expected agent relatedness as a function of distance. 

With each turn, agents may move to any unoccupied location within their Moore neighborhood with equal probability, including their current location. The probability of moving to any of eight neighboring cells within the turn is then, 
\begin{equation}
p_{\text{move}}
=
\sum_{m=0}^{8}
\binom{8}{m}(1-\alpha)^m\alpha^{8-m}
\frac{m}{m+1}.
\end{equation}
Paired agents move twice within a time step, once for each agent. As the lifetime used in this work, $T = 10$, ensures a regular turnover of $10-20 \%$ of the population (pairs are removed when the first dies), individual agents generally pair with neighbors within the first time step. We instead focus on the characteristic migration of agent pairs. 

The diffusion constant, $D$ for a stochastic process, $\textbf{X}(t)$ in two spatial dimensions is defined by, 
\begin{equation}
D \equiv \lim_{t \to \infty}
\frac{1}{4\,t}
\mathbb{E}\!\left[|\mathbf{X}(t)-\mathbf{X}(0)|^2\right].
\end{equation}
This is the mean-squared displacement at large times. Given our walk through Moore neighborhoods, there two options for horizontal and vertical steps, and four options for diagonal steps. Using a Euclidean distance metric, the mean-squared horizontal displacement and mean-squared vertical displacement for paired agents (conditional on motion), 
\begin{equation}
    \mathbb{E}\!\left[|\mathbf{X}(t)|^2\right] = \mathbb{E}[X(t)^2] + \mathbb{E}[Y(t)^2] = 2\cdot \dfrac{3}{2} p_{\textrm{move}}.
\end{equation}
We have assumed the two potential steps taken by a pair are independent, as agents migrate in randomized order. Our diffusion constant per time step is then, 
\begin{equation}
D_{\text{eff}} =
\begin{cases}
\dfrac{3}{8}\,p_{\mathrm{move}}, & \text{individual agent}, \\[8pt]
\dfrac{3}{4}\,p_{\mathrm{move}}, & \text{paired agent}.
\end{cases}
\end{equation}

The displacement distribution in Euclidean, radial coordinates,
\begin{equation}
P(d_E \mid t) = \frac{d_E}{2 D_{\text{eff}} t} \exp\!\left(-\frac{d_E^2}{4 D_{\text{eff}} t}\right).
\end{equation}
This result includes the mean-field impact of occupied neighboring cells, or jamming \citep{vanKampen2007}. As agents have finite lifetimes and are survived by children, the mean-squared displacement from an earliest ancestor ($t \sim 0$), can be recast in terms of the number of generations, $k$, and a mean reproduction time or replacement time per generation, $T_{r}$. We then have, $t = kT_{r}$. Similarly, if an $i$th and j$th$ agent share a most-recent common ancestor, $c$, corresponding to generational distances $g_{i}(c)$ and $g_{j}(c)$, the shortest generational path distance connecting these two agents is $k_{c} = g_{i} + g_{j} $. The effective lived time walked by this branched lineage is $t_\mathrm{lineage} = k_{c}T_{r}$, noting the additive variances of separate walks. 

As our agents exist on a finite, toroidal map, distant branches of lineage will meet or coalesce after a generational timescale, $T_{c}$. The probability of non-coalescence is an exponential, decreasing the expected contribution of agents with larger generational distance, 
\begin{equation}
    P_{\textrm{coal}}(k) \sim \exp  \bigg( -\dfrac{k}{T_{c}} \bigg).
\end{equation}

Owing to motion, the contribution from a generational depth $k$ for agents displaced by a Euclidean distance $d_{E}$ is weighted by diffusion, 
\begin{equation}
     \omega(k;d_{E}) \sim \frac{d_E}{2 D_{\text{eff}} kT_{r}} \exp\!\left(-\frac{d_E^2}{4 D_{\text{eff}} kT_{r}}\right). 
\end{equation}

Given our toy model of relatedness, $r$, described within the text, the expectation for $r$ between two agents at a specific distance, $d_{E}$, is, 
\begin{align}
    \mathbb{E}[r \mid d_{E}] \approx& \sum_{k = 1}^{\infty} \bigg( \dfrac{1}{2} \bigg)^{k} \omega(k;d_{E})P_{\textrm{coal}}(k), \\
    \approx& \sum_{k = 1}^{\infty} \bigg( \dfrac{1}{2} \bigg)^{k}   \frac{d_E}{2 D_{\text{eff}} kT_{r}} \exp\!\left(-\frac{d_E^2}{4 D_{\text{eff}} kT_{r}}\right)    \exp \bigg( -\dfrac{k}{T_{c}} \bigg).
\end{align}
Here, we have denoted a generic generational depth by $k$. The parameters, $T_{r}$, $T_{c}$ and $\alpha$ would be determined through simulation. This is largely an approximation, and does not fully capture the anisotropy of Euclidean-measured distance traversed within Moore neighborhoods, variation in reproduction time or the number of children, variation in $D_{\textrm{eff}}$ owing to the presence of both individual and paired agent populations, and scaling of generational coalescence. Further, populations near spatial capacity are largely jammed, with decreased motion. Instead, walks are subdiffusive or percolative, potentially leading to different functional behavior for the expected relatedness with distance. 

\section{Resource-Independent Population Dynamics}

In our grid of occupancy, $K = n \times n$, we consider the evolving number of individual agents, $N_{\textrm{indiv}}(t)$ and the number of pairs, $N_{\textrm{pair}}(t)$. We first discuss an age-structured, discrete-time dynamical map, yielding expectations that most closely reflect the update scheme used in our agent based model. Tracking of age structure allows for a Markovian mapping. We then assume a continuous-time, well-mixed population described by mean-field behavior, providing an approximate ODE model.

\subsection{Discrete-Time Mapping}

The total population size is, 
\begin{equation}
    N(t) = N_{\textrm{indiv}}(t) + 2 N_{\textrm{pair}}(t).
\end{equation}
The number of individual agents may be further decomposed as a sum of age class populations, $a = 0,1,\dots,T-1$, 
\begin{equation}
    N_{\textrm{indiv}}(t) = \sum_{a = 0}^{T - 1}N_{\textrm{indiv,a}}(t).
\end{equation}
Similarly, the number of agent pairs with a greatest age of $a$ is, 
\begin{equation}
    N_{\textrm{pair}}(t) = \sum_{a = 0}^{T - 1}N_{\textrm{pair,a}}(t).
\end{equation}
The fraction of occupied sites is, 
\begin{equation}
   \alpha(t) =  \dfrac{N_{\textrm{indiv}}(t) + N_{\textrm{pair}}(t)}{K}. 
\end{equation}
If $N_{\textrm{indiv}}(t)$ and $N_{\textrm{pair}}(t)$ are evaluated at the start of the time step, $\alpha(t)$ must be updated after each density-effecting operation within the turn – pairing, reproduction and mortality. 

With each time step, the expected number of individuals per cell at a given time is, $N_{\textrm{indiv}}(t)/K$, corresponding to a well-mixed population. As agents may check all eight Moore neighbors for a potential match, requiring only a single other individual, the probability of matching is, 
\begin{equation}
    p_{\text{match}}(t) = 1 - \bigg(1 - \dfrac{N_{\textrm{indiv}}(t)}{K} \bigg)^{8}.
\end{equation}
Alternatively, $p_{\text{match}}(t)$ can be expressed as a function of each age class, 
\begin{equation}
    p_{\text{match, a}}(t) =  1 - \bigg(1 - \dfrac{N_{\textrm{indiv, a}}(t)}{K} \bigg)^{8}.
\end{equation}

Within our numerical simulation, with each new pair, two individual agents are removed from the pool, lowering $N_{\textrm{indiv}}$. To account for sequential depletion, we can treat pairing as a random matching process, where the expected number of new pairs from the whole population within the time step is \citep{Diekmann2013}, 
\begin{equation}
\mathbb{E}\!\left[\Delta N_{\mathrm{pair}}^{(+)}(t)\right]
=
\frac{N_{\mathrm{indiv}}(t)}{2}
\left[
1 - \exp\!\left(
- p_{\text{match}}(t)
\right)
\right].
\end{equation}
Here, $N_{\mathrm{indiv}}(t)$ would be evaluated at the start of the time step. To determine the number of new pairs of maximum age, $a$, we must account for the distribution of ages within the population. We define the probability, 
\begin{equation}
    \pi_{a}(t) = \mathbb{P}(\text{max}(A,B) = a) = \dfrac{N_{\textrm{indiv, a}}(t) \big(N_{\textrm{indiv}}(t) - \sum_{b > a}  N_{\textrm{indiv, b}}(t) \big)}{ N_{\textrm{indiv}}(t)^{2}}.
\end{equation}
Given an individual of age $a$, $\pi_{a}(t)$ is the probability that another agent has an age less than or equal to $a$. The expected number of new pairs of each age class is then, 
\begin{equation}
\mathbb{E}\!\left[\Delta N_{\mathrm{pair},a}^{(+)}(t)\right]
=
\pi_a(t)\,
\mathbb{E}\!\left[\Delta N_{\mathrm{pair}}^{(+)}(t)\right].
\end{equation}

Each agent pair then has the opportunity to reproduce, dependent on the existence of a single neighboring, empty space on the map. The initial probability follows from the reduction in density after pairing, 
\begin{equation}
    p_{\text{empty}} = 1 - \left(1 - \alpha + \dfrac{\mathbb{E}\!\left[\Delta N_{\mathrm{pair}}^{(+)}(t)\right]}{K} \right)^{8}.
\end{equation}
Reflecting the diminished number of free spaces with each child, the expected number of newly placed children is, 
\begin{equation}
\mathbb{E}\!\left[\Delta N^{(+)}_{\mathrm{indiv}}(t)\right]
= \left[ N_{\mathrm{pair}}(t) + \mathbb{E}\!\left[\Delta N^{(+)}_{\mathrm{pair}}(t)\right] \right] \left[ 1 - \exp \left( - p_{\text{empty}}(t) \right)  \right]. 
\end{equation}

At the end of each turn, individuals and pairs' ages are incremented, transitioning populations to the next age class. Individuals who are born within the turn are excluded from this update, but counted in the overall population size. Agents and pairs with a greatest age of $T$ are killed. A negligible number of individuals are expected to live until $T$ unpaired, with probability $\sim$$(1-p_{\textrm{match}})^{T}$. At late, well-mixed times, we expect roughly a fraction, $1/T$, of all pairs to be removed. The final discrete-time mapping is described by, 
\begin{equation}
\left\{
\begin{aligned}
N_{\mathrm{indiv}}(t+1)
&=
\sum_{a=0}^{T-2} N_{\mathrm{indiv,a}}(t)
- 2\,\mathbb{E} \left[\Delta N_{\mathrm{pair}}^{(+)}(t) \right]
+ \mathbb{E}\left[\Delta N_{\mathrm{indiv}}^{(+)}(t)\right],\\[1em]
N_{\mathrm{pair}}(t+1) 
&= 
\sum_{a=0}^{T-2} N_{\mathrm{pair,a}}(t) 
+ \sum_{a=0}^{T-2} \mathbb{E}\left[\Delta N_{\mathrm{pair},a}^{(+)}(t)\right].
\end{aligned}
\right.
\end{equation}
\begin{equation}
\begin{aligned}
N_{\mathrm{indiv,}0}(t+1) &= \mathbb{E}\left[\Delta N_{\mathrm{indiv}}^{(+)}(t)\right], \\
N_{\mathrm{indiv,}a+1}(t+1) &= N_{\mathrm{indiv,}a}^{\star}(t),
\\
N_{\mathrm{pair,}a+1}(t+1) &= N_{\mathrm{pair,}a}^{\star}(t),
\qquad a=0,\dots,T-2.
\end{aligned}
\end{equation}
Here, the superscript, $^{*}$, denotes the post-interaction population prior to aging. Ages are then incremented, forming classes of ages, $a = 0, \dots, T - 1$, for the next time step.

\subsection{Mean-Field ODE Description}

We also discuss a phenomenological ODE description of the resource-independent system, assuming a well-mixed population.  Instantaneously, the expression for probabilistic matching remains the same, $p_{\text{match}}$. The instantaneous rate of pair formation, 
\begin{equation}
\frac{dN_{\mathrm{pair}}^{(+)}}{dt}
=
\frac{1}{2}\,
N_{\mathrm{indiv}}(t)
\left[
1 - \left(1 - \frac{N_{\mathrm{indiv}}(t)}{K}\right)^8
\right].
\end{equation}
While pairs are removed in simulation when the first agent reaches $T$, if the time to matching is relatively small relative to the total lifespan, we can approximate the lifespan of agent pairs as $T$. The total growth of pairs is then expressed as, 
\begin{equation}
    \frac{dN_{\mathrm{pair}}}{dt}
=
\frac{1}{2}\,
N_{\mathrm{indiv}}(t)
\left[
1 - \left(1 - \frac{N_{\mathrm{indiv}}(t)}{K}\right)^8
\right] - \dfrac{N_{\text{pair}}(t)}{T}.
\end{equation}

Change in the individual agent number is a function of both those lost to the paired population, those lost through the death, and increase through reproduction. The probability of an agent existing for $T$ time steps unpaired is approximately $(1 - p_{\textrm{match}})^{T}$, negligible for parameter values of interest. Assuming the matching process is fast relative to $T$, we drop loss through death in our expectation,

\begin{align}
\frac{dN_{\mathrm{indiv}}}{dt}
=&
- N_{\mathrm{indiv}}(t)
\left[
1 - \left(1 - \frac{N_{\mathrm{indiv}}(t)}{K}\right)^8
\right] \\
&\quad
+ N_{\mathrm{pair}}(t)
\Big[1-(1-\alpha(t))^8\Big].
\end{align}

The result is an expression for the rate of change in the number of individuals through pairing and spatially-limited reproduction. 

An extinction state is possible as well as a stable positive-density population equilibrium. 

\section{Resource-Dependent Population Dynamics}

Without resource redistribution, survival before natural death depends only on $\mu$ and $\phi$, such that the probability of survival, 
\begin{equation}
    P_{\textrm{surv}}(\mu, \phi) = 1 - e^{-\mu} \sum_{i = 0}^{\phi - 1} \dfrac{\mu^{i}}{i!}.
\end{equation}
Here, $0 \leq P_{\textrm{surv}}(\mu, \phi) \leq 1$. This probability scales the size of the surviving population at each time step. Offspring born within the timestep are not dependent on resources. In the case of pairs, each agent must survive independently, leading to a quadratic resource dependence. 

\subsection{Discrete-Time Mapping}

As reproduction occurs after pairing and survival, the expected number of newly placed individuals within the turn is scaled by $P_{\mathrm{surv}}(\mu,\phi)^{2}$. Additionally, the initial expected number of neighboring free spaces available also reflects those lost to resource death, 

\begin{align}
p_{\text{empty}}
&=
1 -
\left(
1 -
\frac{
P_{\mathrm{surv}}(\mu,\phi)
\left[
N_{\mathrm{indiv}}(t)
- 2\,\mathbb{E}\!\left[\Delta N_{\mathrm{pair}}^{(+)}(t)\right]
\right]
}{
K
}
\right.
\nonumber\\
&\qquad
\left.
-
\frac{
P_{\mathrm{surv}}(\mu,\phi)^{2}
\left[
N_{\mathrm{pair}}(t)
+
\mathbb{E}\!\left[\Delta N_{\mathrm{pair}}^{(+)}(t)\right]
\right]
}{
K
}
\right)^{8}.
\end{align}

Our discrete-time mapping is now, 
\begin{align}
N_{\mathrm{indiv}}(t+1)
&=
P_{\mathrm{surv}}(\mu,\phi)
\left[
\sum_{a=0}^{T-2} N_{\mathrm{indiv},a}(t)
\right]
\nonumber\\
&\quad
- 2\,\mathbb{E}\!\left[\Delta N_{\mathrm{pair}}^{(+)}(t)\right]
+ P_{\mathrm{surv}}(\mu,\phi)^{2}
\,\mathbb{E}\!\left[\Delta N_{\mathrm{indiv}}^{(+)}(t)\right],
\\[1em]
N_{\mathrm{pair}}(t+1)
&=
P_{\mathrm{surv}}(\mu,\phi)^{2}
\left[
\sum_{a=0}^{T-2} N_{\mathrm{pair},a}(t)
+
\sum_{a=0}^{T-2}
\mathbb{E}\!\left[\Delta N_{\mathrm{pair},a}^{(+)}(t)\right]
\right].
\end{align}

\begin{equation}
\begin{aligned}
N_{\mathrm{indiv},0}(t+1)
&=
\mathbb{E}\!\left[\Delta N_{\mathrm{indiv}}^{(+)}(t)\right],
\\
N_{\mathrm{indiv},a+1}(t+1)
&=
N_{\mathrm{indiv},a}^{\star}(t),
\\
N_{\mathrm{pair},a+1}(t+1)
&=
N_{\mathrm{pair},a}^{\star}(t),
\qquad a = 0,\dots,T-2.
\end{aligned}
\end{equation}
$N^{*}$ again represents the post-interaction population. Ages are then incremented, forming classes of ages, $a = 0, \dots, T - 1$, for the next time step.

\subsection{Mean-Field ODE Description}

In the well-mixed, continuous time limit, we also provide an approximate, phenomenological differential equation system representative of the coupling between pair and individual populations. We assume that pair death occurs at a rate, $1/T$, and that all individual agents successfully pair before reaching an age of $T$. Both populations experience resource hazard.  

\begin{align}
\frac{dN_{\mathrm{indiv}}}{dt}
&=
-
N_{\mathrm{indiv}}(t)
\left[
1-
\left(
1-\frac{N_{\mathrm{indiv}}(t)}{K}
\right)^8
\right]
\nonumber\\
&\quad
+
P_{\mathrm{surv}}(\mu,\phi)^2\,
N_{\mathrm{pair}}(t)
\left[
1-
\left(
1-\alpha(t)
\right)^8
\right]
\nonumber\\
&\quad
-
\bigl(1-P_{\mathrm{surv}}(\mu,\phi)\bigr)
N_{\mathrm{indiv}}(t),
\\[1.2em]
\frac{dN_{\mathrm{pair}}}{dt}
&=
\frac{1}{2}\,
N_{\mathrm{indiv}}(t)
\left[
1-
\left(
1-\frac{N_{\mathrm{indiv}}(t)}{K}
\right)^8
\right]
\nonumber\\
&\quad
-
\left(
\frac{1}{T}
+
1-P_{\mathrm{surv}}(\mu,\phi)^2
\right)
N_{\mathrm{pair}}(t).
\end{align}

$P_{\mathrm{surv}}(\mu,\phi)$ acts as a bifurcation parameter, separating an extinction state from a stable positive-density equilibrium population. Values near $P_{\textrm{critical}}(\mu,\phi)$ may appear metastable, owing to transient low-density, finite-size population fluctuations prior to extinction. As $P_{\mathrm{surv}}(\mu,\phi)$ and $\mu$ increases, the equilibrium population size is increasingly limited by spatial saturation. 

\section{Redistribution-Dependent Population Dynamics}
\label{app:redist}

\begin{figure}[t]
    \centering
    \includegraphics[width=0.75\linewidth]{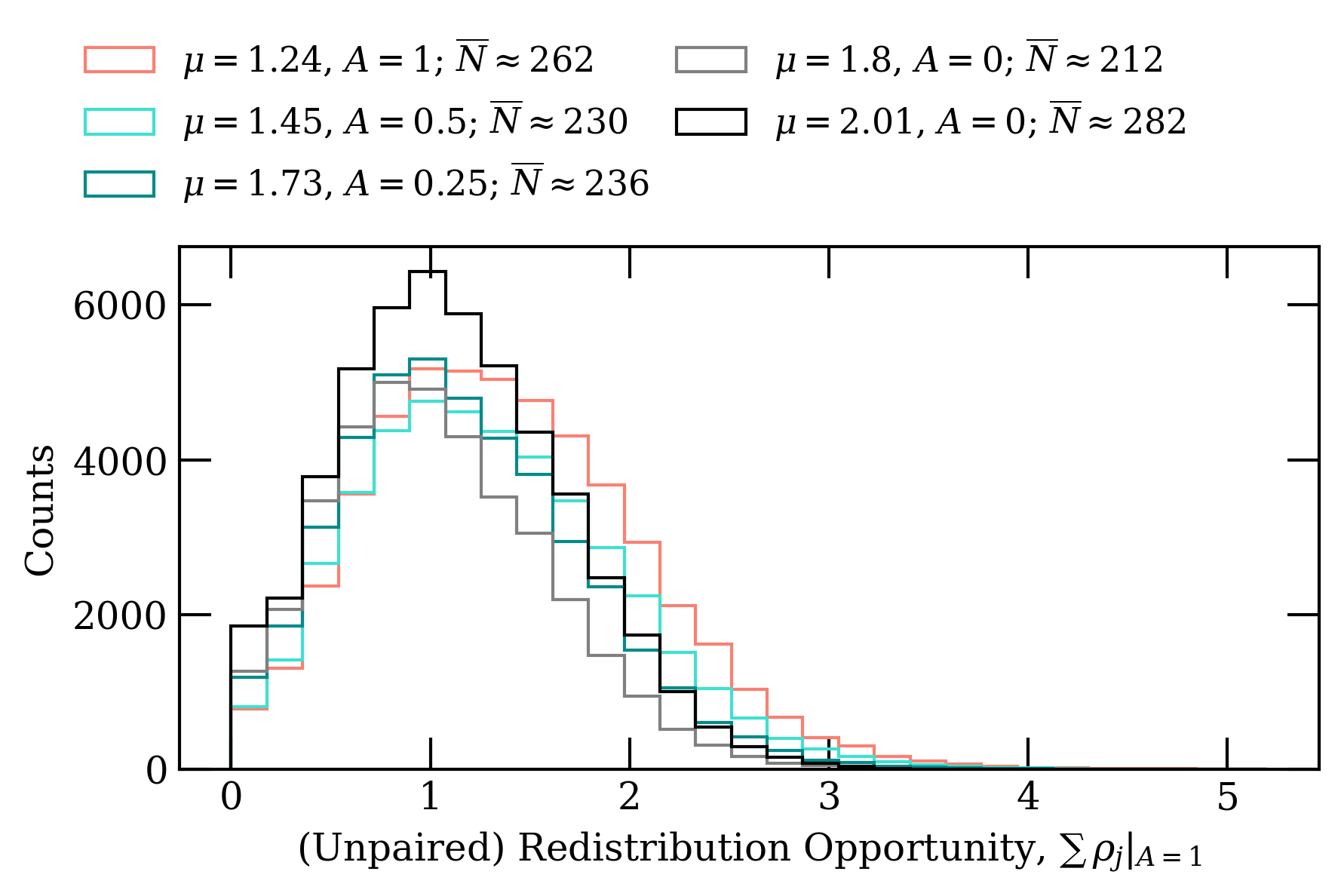}
    \caption{Redistribution opportunity excluding the contribution of the pair partner. We provide distributions of summed agent redistribution opportunity excluding the transfer fraction associated with the partners of paired agents, $\sum_{j} \rho_{j}|_{A = 1}$. Distributions reflect statistically independent samples from ensembles of 200 simulations. We evaluate these sums with the scale $A = 1$. The resulting distributions reflect agent extended generational and spatial kinship proximity. Dependence on redistribution leads to populations with greater mean agent node strength or relatedness. The distribution of node strengths also shows a heavier high-value tail, suggesting greater organization or potential increased network centrality. }
    \label{fig:no_pair}
\end{figure}

We provide an additional figure characterizing the summed redistribution opportunities of our populations. The value, $\sum_{j} \rho_{j}$, generally traces extended kinship network structure – both the distribution of generational and spatial closeness. However, we included a slightly stylized or artificial definition for the transfer fraction between paired individuals. After comparing distributions of $\sum_{j} \rho_{j}$ for different populations, we found that the most redistribution-dependent populations consistently achieve the highest $\sum \rho_{j}$ values, and generally show higher mean $\sum_{j} \rho_{j}$. To ensure that this was not an effect of differences in agent pairing, and truly dependent on the extended network structure, we have also plotted $\sum_{j} \rho_{j}$ with the contribution from the pair partner removed from the sum in Figure \ref{fig:no_pair}. We find that our original qualitative result does not depend on the redistribution between partners, and instead reflects a genuine change in the extended network topology. 

\section*{Acknowledgments}
This work was supported in part through computational resources and services provided by the Institute for Cyber-Enabled Research at Michigan State University.

\section*{Declaration of Generative AI and AI-Assisted Technologies in the Manuscript Preparation Process}
During the preparation of this work the author used ChatGPT for the purpose of literature recommendations, literature summaries supplemental to the standard research process, and suggestions on text readability in selected sections. The author takes full responsibility for the content and concept of the published article.

\bibliographystyle{elsarticle-harv}
\bibliography{notes}

\end{document}